# A Review on the Use of Blockchain for the Internet of Things

**TIAGO M. FERNÁNDEZ-CARAMÉS[1], (Senior Member, IEEE),
AND PAULA FRAGA-LAMAS[1], (Member, IEEE)**
[1]Department of Computer Engineering, Faculty of Computer Science, Campus de Elviña s/n, Universidade da Coruña, 15071, A Coruña, Spain.
(e-mail: tiago.fernandez@udc.es; paula.fraga@udc.es)

Corresponding authors: Tiago M. Fernández-Caramés and Paula Fraga-Lamas (e-mail: tiago.fernandez@udc.es; paula.fraga@udc.es).

This work was supported by the Xunta de Galicia (ED431C 2016-045, ED341D R2016/012, ED431G/01), the Agencia Estatal de Investigación of Spain (TEC2015-69648-REDC, TEC2016-75067-C4-1-R) and ERDF funds of the EU (AEI/FEDER, UE). Paula Fraga-Lamas would also like to thank the support of BBVA and the BritishSpanish Society Grant.

**ABSTRACT** The paradigm of Internet of Things (IoT) is paving the way for a world where many of our daily objects will be interconnected and will interact with their environment in order to collect information and automate certain tasks. Such a vision requires, among other things, seamless authentication, data privacy, security, robustness against attacks, easy deployment and self-maintenance. Such features can be brought by blockchain, a technology born with a cryptocurrency called Bitcoin. In this paper it is presented a thorough review on how to adapt blockchain to the specific needs of IoT in order to develop Blockchain-based IoT (BIoT) applications. After describing the basics of blockchain, the most relevant BIoT applications are described with the objective of emphasizing how blockchain can impact traditional cloud-centered IoT applications. Then, the current challenges and possible optimizations are detailed regarding many aspects that affect the design, development and deployment of a BIoT application. Finally, some recommendations are enumerated with the aim of guiding future BIoT researchers and developers on some of the issues that will have to be tackled before deploying the next generation of BIoT applications.

## I. INTRODUCTION

The Internet of Things (IoT) is expanding at a fast pace and some reports [1] predict that IoT devices will grow to 26 billions by 2020, which are 30 times the estimated number of devices deployed in 2009 and is far more than the 7.3 billion smartphones, tablets and PCs that are expected to be in use by 2020. Moreover, some forecasts [2] anticipate a fourfold growth in Machine-to-Machine (M2M) connections in the next years (from 780 million in 2016 to 3.3 billion by 2021), which may be related to a broad spectrum of applications like home automation [3], transportation [4], defense and public safety [5], wearables [6] or augmented reality [7], [8].

In order to reach such a huge growth, it is necessary to build an IoT stack, standardize protocols and create the proper layers for an architecture that will provide services to IoT devices. Currently, most IoT solutions rely on the centralized server-client paradigm, connecting to cloud servers through the Internet. Although this solution may work properly nowadays, the expected growth suggests that new paradigms will have to be proposed. Among such proposals, decentralized architectures were suggested in the past to create large Peer-to-Peer (P2P) Wireless Sensor Networks (WSNs) [9]–[11], but some pieces were missing in relation to privacy and security until the arrival of blockchain technology. Therefore, as it is illustrated in Figure 1, in the last years pre-IoT closed and centralized mainframe architectures evolved towards IoT open-access cloud-centered alternatives, being the next step the distribution of the cloud functionality among multiple peers, where blockchain technology can help.

Blockchain technologies are able to track, coordinate, carry out transactions and store information from a large amount of devices, enabling the creation of applications that require no centralized cloud. Some companies like IBM go further and talk about blockchain as a technology for de-

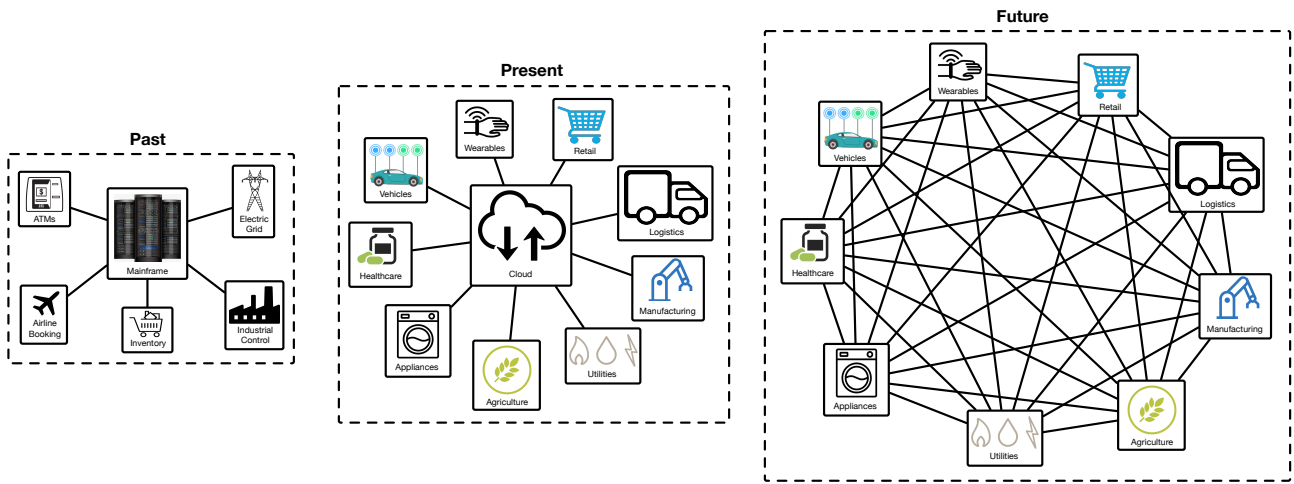

FIGURE 1: Past, present and future IoT architectures.

mocratizing the future IoT [12], since it addresses the current critical challenges for its massive adoption:

- Many IoT solutions are still expensive due to costs related to the deployment and maintenance of centralized clouds and server farms. When such an infrastructure is not created by the supplier, the cost comes from middlemen.
- Maintenance is also a problem when having to distribute regular software updates to millions of smart devices.
- After Edward Snowden leaks [13], [14], it is difficult for IoT adopters to trust technological partners who, in general, give device access and control to certain authorities (i.e., governments, manufacturers or service providers), allowing them to collect and analyze user data. Therefore, privacy and anonymity should be at the core of future IoT solutions.
- Lack of trust is also fostered by closed-source code. To increase trust and security, transparency is essential, so open-source approaches should be taken into account when developing the next generation of IoT solutions. It is important to note that open-source code, like closed-source code, is still susceptible to bugs and exploits, but, since it can be monitored constantly by many users, it is less prone to malicious modifications from third parties.

Blockchain technology has been growing at an astounding pace over the past two years. As reported by Statista [15], investments by venture capitalists in blockchain start-ups rose from 93 million to 550 million U.S. dollars from 2013 to 2016. Furthermore, the market for blockchain technology worldwide is forecast to grow to 2.3 billion U.S. dollars by 2021. According to McKinsey & Company, although it is still in a nascent stage, blockchain technology may reach its full potential within the next 4 years based on its current pace of evolution [16]. In addition, as of writing, there are over 1,563 digital coins [17], just a few years after Bitcoin [18], the cryptocurrency that originated the blockchain, was born.

Bitcoin is a digital coin whose transactions are exchanged in a decentralized trustless way combining peer-to-peer file sharing with public-key cryptography. Public keys are alphanumeric strings formed by 27 to 32 characters that are used to send and receive Bitcoins, avoiding the necessity of making use of personal information to identify users. One feature that characterizes Bitcoin is miners, who receive coins for their computational work to verify and store payments in the blockchain. Such payments, like in any other currency, are performed in exchange of products, services or fiat money. This paper is not aimed at detailing the inner workings of Bitcoin, but the interested readers can find good overviews on how Bitcoin works in [19]–[21].

The use of cryptocurrencies based on blockchain technology is said to revolutionize payments thanks to their advantages respect to traditional currencies. Since middlemen are removed, merchant payment fees can be reduced below 1 % and users do not have to wait days for transfers, receiving funds immediately. Modern cryptocurrencies can be divided into three elements [19]: blockchain, protocol and currency. It must be indicated that a coin can implement its own currency and protocol, but its blockchain may run on the blockchain of another coin like Bitcoin or Ethereum [22]. For instance, Counterparty [23] has its own currency and protocol, but it runs on the Bitcoin blockchain.

In the case of a cryptocurrency, the blockchain acts as a ledger that stores all the coin transactions that have been performed. This means that the blockchain grows continuously, adding new blocks every certain time intervals. A full node (a computer that validates transactions) owns a copy of the whole blockchain, which also contains information about user addresses and balances. If the blockchain is public, in can be queried through a block explorer like Blockchain.info in order to obtain the transactions related to a specific address.

Therefore, the key contribution of blockchain is that it provides a way to carry out transactions with another person or entity without having to rely on third-parties. This is possible thanks to many decentralized miners (i.e., accountants) that scrutinize and validate every transaction. This contribution allowed the Bitcoin blockchain to provide a solution to the Byzantine Generals' Problem [24], since it is able to reach an agreement about something (a battle plan) among multiple parties (generals) that do not trust each other, when only exchanging messages, which may come from malicious third-parties (traitors) that may try to mislead them. In the case of cryptocurrencies, this computational problem is related to the double-spend problem, which deals with how to confirm that some amount of digital cash was not already spent without the validation of a trusted third-party (i.e., usually, a bank) that keeps a record of all the transactions and user balances.

IoT shares some common problems with cryptocurrencies, since in an IoT system there are many entities (nodes, gateways, users) that do not necessarily trust each other when performing transactions. However, there are several aspects that differentiate IoT from digital currencies, like the amount of computing power available in the nodes or the necessity for minimizing the energy consumed in devices powered with batteries. Therefore, this paper studies such similarities and analyzes the advantages that blockchain can bring to IoT despite its current practical limitations. Moreover, the main Blockchain-based IoT (BIoT) architectures and improvements that have already proposed are reviewed. Furthermore, the most relevant future challenges for the application of blockchain to IoT are detailed.

Other authors have previously presented surveys on the application of blockchain to different fields. For instance, in [25] it is provided an extensive description on the basics of blockchain and smart contracts, and it is given a good overview on the application and deployment of BIoT solutions. However, although the paper provides very useful information, it does not go deep into the characteristics of the ideal BIoT architecture or on the possible optimizations to be performed for creating BIoT applications. Another interesting work is presented in [26], where the authors provide a generic review on the architecture and the different mechanisms involved in blockchain, although it is not focused on its application to IoT. Similarly, in [27] and [28] different researchers give overviews on blockchain, but they emphasize its application to different Big Data areas and multiple industrial applications. Finally, it is worth mentioning the systematic reviews presented in [29] and [30], which analyze the sort of topics that papers in the literature deal with when proposing the use of blockchain.

Unlike the reviews previously mentioned, this work presents a holistic approach to blockchain for IoT scenarios, including not only the basics on blockchain-based IoT applications, but also a thorough analysis on the most relevant aspects involved on their development, deployment and optimization. It is also the aim of this work to envision the potential contribution of blockchain for revolutionizing the IoT industry and confront today challenges.

The remainder of this paper is organized as follows. Section II describes the basics of blockchain technologies: how they work, which types exist and how to decide if it is appropriate to make use of a blockchain. Section III presents the most relevant BIoT applications. Section IV reviews critical aspects to be optimized in a blockchain in order to adapt it to an IoT application. Section V describes the main shortcomings of current BIoT applications and outlines the primary technical challenges they face. Section VI identifies further medium-term challenges and proposes recommendations for IoT developers. Finally, Section VII is devoted to conclusions.

## II. BLOCKCHAIN BASICS

A blockchain is like a distributed ledger whose data are shared among a network of peers. As it was previously mentioned, it is considered as the main contribution of Bitcoin, since it solved a longer-lasting financial problem known as the double-spend problem. The solution proposed by Bitcoin consisted in looking for the consensus of most mining nodes, who append the valid transactions to the blockchain.

Although the concept of blockchain was originated as a tool for a cryptocurrency, it is not necessary to develop a cryptocurrency to use a blockchain and build decentralized applications [31]. A blockchain, as its name implies, is a chain of timestamped blocks that are linked by cryptographic hashes. To introduce the reader into the inner workings of a blockchain, the next subsections describe its basic characteristics and functioning.

### A. BLOCKCHAIN BASIC FUNCTIONING

In order to use a blockchain, it is first required to create a P2P network with all the nodes interested in making use of such a blockchain. Every node of the network receives two keys: a public key, which is used by the other users for encrypting the messages sent to a node, and a private key, which allows a node to read such messages. Therefore, two different keys are used, one for encrypting and another for decrypting. In practice, the private key is used for signing blockchain transactions (i.e., to approve such transactions), while the public key works like a unique address. Only the user with the proper private key is able to decrypt the messages encrypted with the corresponding public key. This is called asymmetric cryptography. A detailed explanation of its inner workings is out of the scope of this paper, but the interested reader can obtain further details in [32], [33].

When a node carries out a transaction, it signs it and then broadcasts it to its one-hop peers. The fact of signing the transaction in a unique way (using the private key) enables authenticating it (only the user with a specific private key can sign it) and guarantees integrity (if there is an error during the transmission of the data, it will not be decrypted). As the peers of the node that broadcasts the transaction receive the signed transaction, they verify that it is valid before retransmitting it to other peers, thus, contributing to its spread through the network. The transactions disseminated in this way and that are considered valid by the network are ordered and packed into a timestamped block by special nodes called miners. The election of the miners and the data included into the block depend on a consensus algorithm (a more detailed definition of the concept of consensus algorithm is given later in Section IV-D).

The blocks packed by a miner are then broadcast back into the network. Then the blockchain nodes verify that the broadcast block contains valid transactions and that it references the previous block of the chain by using the corresponding hash. If such conditions are not fulfilled, the block is discarded. However, if both conditions are verified successfully, the nodes add the block to their chain, updating the transactions.

### B. TYPES OF BLOCKCHAINS

There are different types of blockchains depending on the managed data, on the availability of such data, and on what actions can be performed by a user. Thus, it can be distinguished between public and private, and permissioned and permissionless blockchains.

It is important to indicate that some authors use the terms public/permissionless and private/permissioned as synonyms, what may be coherent when talking about cryptocurrencies, but that is not the case for IoT applications, where it is important to distinguish between authentication (who can access the blockchain; private versus public) and authorization (what an IoT device can do; permissionless versus permissioned). Nonetheless, note that such distinctions are still in debate and the definitions given next might differ from others in the literature.

In public blockchains anyone can join the blockchain without the approval of third-parties, being able to act as a simple node or as miner/validator. Miners/validators are usually given economic incentives in public blockchains like Bitcoin, Ethereum or Litecoin [34].

In the case of private blockchains, the owner restricts network access. Many private blockchains are also permissioned in order to control which users can perform transactions, carry out smart contracts (a concept defined later in Section III) or act as miners in the network, but note that not all private blockchains are necessarily permissioned. For instance, an organization can deploy a private blockchain based on Ethereum, which is permissionless. Examples of permissioned blockchains are the ones used by Hyperledger-Fabric [35] or Ripple [36].

It can also be distinguished between blockchains aimed exclusively at tracking digital assets (e.g., Bitcoin) and blockchains that enable running certain logic (i.e., smart contracts). Moreover, there are systems that make use of tokens (e.g., Ripple), while others do not (e.g., Hyperledger). Note that such tokens are not necessarily related to the existence of a cryptocurrency, but they may be used as internal receipts that prove that certain events happened at certain time instants.

As a summary, the different types of blockchains are depicted in Figure 2 together with several examples of implementations.

### C. DETERMINING THE NEED FOR USING A BLOCKCHAIN

Before delving into the details on how to make use of a blockchain for IoT applications, it must be first emphasized that a blockchain is not always the best solution for every IoT scenario. Traditional databases or Directed Acyclic Graph (DAG) based ledgers [37] may be a better fit for certain IoT applications. Specifically, in order to determine if the use of a blockchain is appropriate, a developer should decide if the following features are necessary for an IoT application:

- Decentralization. IoT applications demand decentralization when there is not a trusted centralized system. However, many users still trust blindly certain companies, government agencies or banks, so if there is mutual trust, a blockchain is not required.
- P2P exchanges. In IoT most communications go from nodes to gateways that route data to a remote server or cloud. Communications among peers at a node level are actually not very common, except for specific applications, like in intelligent swarms [38] or in mist computing systems [39]. There are also other paradigms that foster communications among nodes at the same level, as it happens in fog computing with local gateways [40], [41].
- Payment system. Some IoT applications may require to perform economic transactions with third parties, but many applications do not. Moreover, economic transactions can still be carried out through traditional payment systems, although they usually imply to pay transaction fees and it is necessary to trust banks or middlemen.
- Public sequential transaction logging. Many IoT networks collect data that need to be timestamped and stored sequentially. Nonetheless, such needs may be easily fulfilled with traditional databases, especially in cases where security is guaranteed or where attacks are rare.
- Robust distributed system. Distributed systems can also be built on top of clouds, server farms or any form of traditional distributed computing systems [42]. The need of this feature is not enough to justify the use of a blockchain: there also has to be at least a lack of trust in the entity that manages the distributed computing system.
- Micro-transaction collection. Some IoT applications [43], [44] may need to keep a record of every transaction to maintain traceability, for auditing purposes or because Big Data techniques will be applied later [45], [46]. In these situations, a sidechain may be useful [47]. However, other applications do not need to store every collected value. For example, in remote agricultural monitoring, where communications are expensive, it is usual to make use of IoT nodes that wake up every hour to obtain environmental data from sensors. In such cases, a local system may collect and store the data, and once a day it transmits the processed information

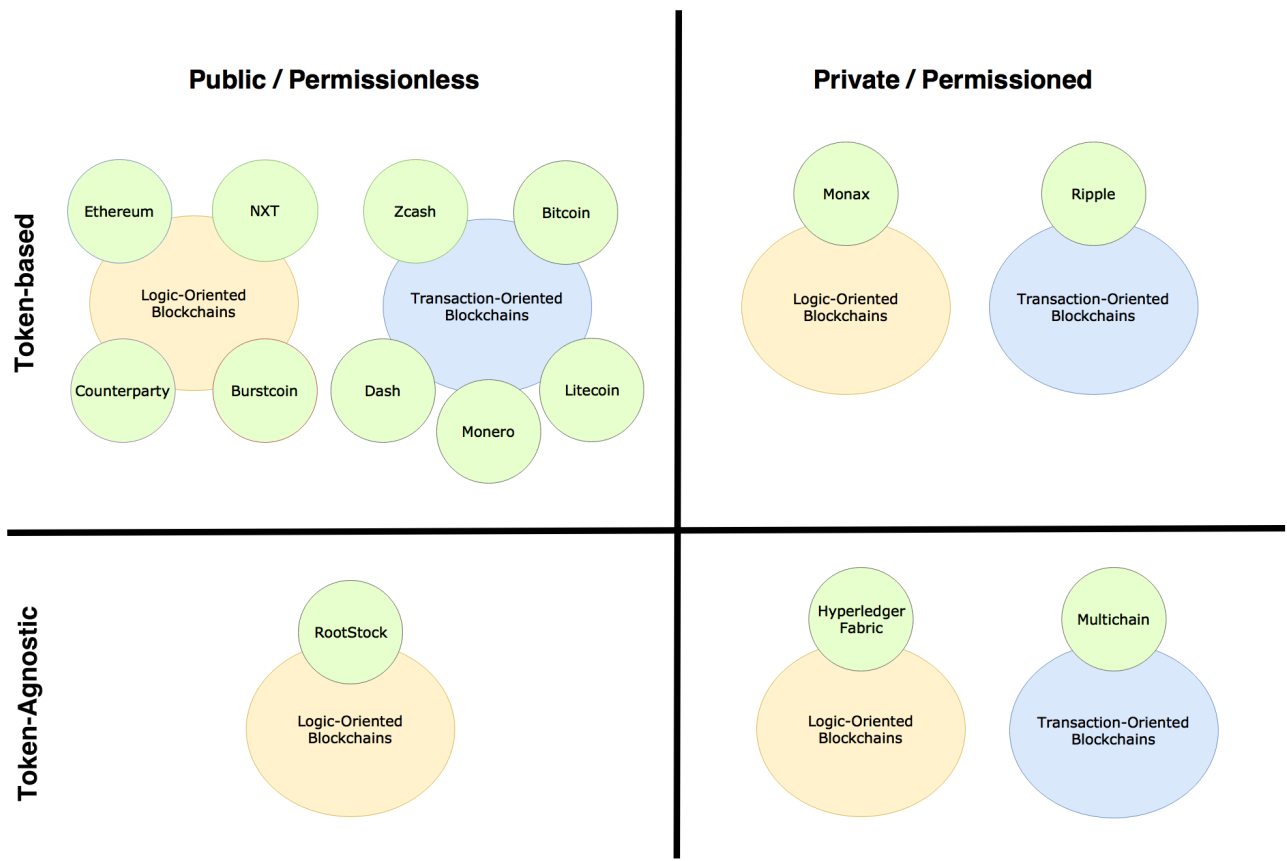

FIGURE 2: Blockchain taxonomy and practical examples.

altogether in one transaction [48].

Figure 3 shows a generic flow diagram that allows for determining the type of blockchain that is necessary depending on the characteristics of an IoT system.

## III. BIOT APPLICATIONS

Blockchain technology can be applied in many fields and use cases. Some authors [19] suggested that blockchain applicability evolution started with Bitcoin (blockchain 1.0), then evolved towards smart contracts (blockchain 2.0) and later moved to justice, efficiency and coordination applications (blockchain 3.0).

Regarding smart contracts, they are defined as pieces of self-sufficient decentralized code that are executed autonomously when certain conditions are met. Smart contracts can be applied in many practical cases, including international transfers, mortgages or crowd funding [49].

Ethereum is arguably the most popular blockchain-based platform for running smart contracts, although it can actually run other distributed applications and interact with more than one blockchain. In fact, Ethereum is characterized by being Turing-complete, which is a mathematical concept that indicates that Ethereum's programming language can be used to simulate any other language. A detailed explanation on how smart contracts work is out of the scope of this paper, but the interested reader can find a really good description in Section II.D of [25].

Beyond cryptocurrencies and smart contracts, blockchain technologies can be applied in different areas (the most relevant are shown in Figure 4) where IoT applications are involved [29], like sensing [50], [51], data storage [52], [53], identity management [54], timestamping services [55], smart living applications [56], intelligent transportation systems [57], wearables [58], supply chain management [59], mobile crowd sensing [60], cyber law [61] and security in mission-critical scenarios [62].

Blockchain can also be used in IoT agricultural applications. For example, in [63] it is presented a traceability system for tracking Chinese agri-food supplies. The system is based on the use of Radio Frequency Identification (RFID) and a blockchain, being its aim to enhance food safety and quality, and to reduce losses in logistics.

Other researchers focused on managing IoT devices through a blockchain [64]. Such researchers proposed a system able to control and configure IoT devices remotely. The system stores public keys in Ethereum while private keys are saved on each IoT device. The authors indicate that the use of Ethereum is essential, since it allows them to write their own code to run on top of the network. Moreover, updating the code on Ethereum modifies the behavior of the IoT devices,

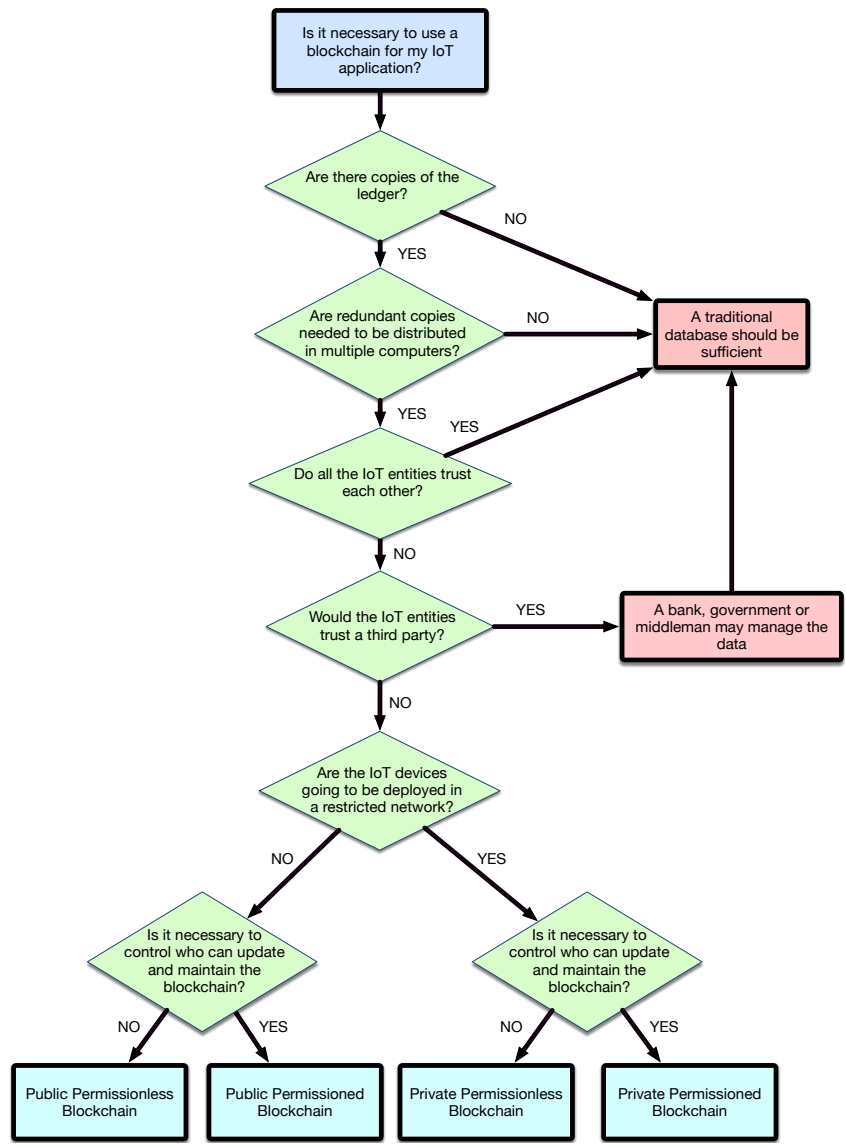

FIGURE 3: Flow diagram for deciding when to use blockchain in an IoT application.

what simplifies maintenance and bug corrections.

The energy sector can also be benefited from the application of a blockchain to IoT or to the Internet of Energy (IoE) [65]–[67]. An example is detailed in [68], where the authors propose a blockchain-based system that allows IoT/IoE devices to pay each other for services without human intervention. In the paper it is described an implementation that shows the potential of the system: a smart cable that connects to a smart socket is able to pay for the electricity consumed. In addition, to reduce the transaction fees of cryptocurrencies like Bitcoin, the researchers present a single-fee micro-payment protocol that aggregates several small payments into a larger transaction.

Healthcare BIoT applications are found in the literature as well. For instance, in [69] it is presented a traceability application that makes use of IoT sensors and blockchain technology to verify data integrity and public accessibility to temperature records in the pharmaceutical supply chain. This verification is critical for the transport of medical products in order to ensure their quality and environmental conditions (i.e., their temperature and relative humidity). Thus, every shipped parcel contains a sensor that transfers the collected data to the blockchain where a smart contract determines whether the received values remain within the allowed range. Another healthcare BIoT application is detailed in [70], where it is presented the architecture of a blockchain-based platform for clinical trials and precision medicine. It is also worth mentioning the work described in [71], which presents a generic smart healthcare system that makes use of IoT devices, cloud and fog computing [72], a blockchain, Tor [73] and message brokers.

IoT low-level security can also be enhanced by blockchain technology. Specifically, it can be improved remote attestation, which is the process that verifies whether the underlying Trusted Computer Base (TCB) of a device is trustworthy [74]. This verification can be performed by managing the TCB measurements obtained by using ARM TrustZone [75] and a blockchain, where they are stored securely.

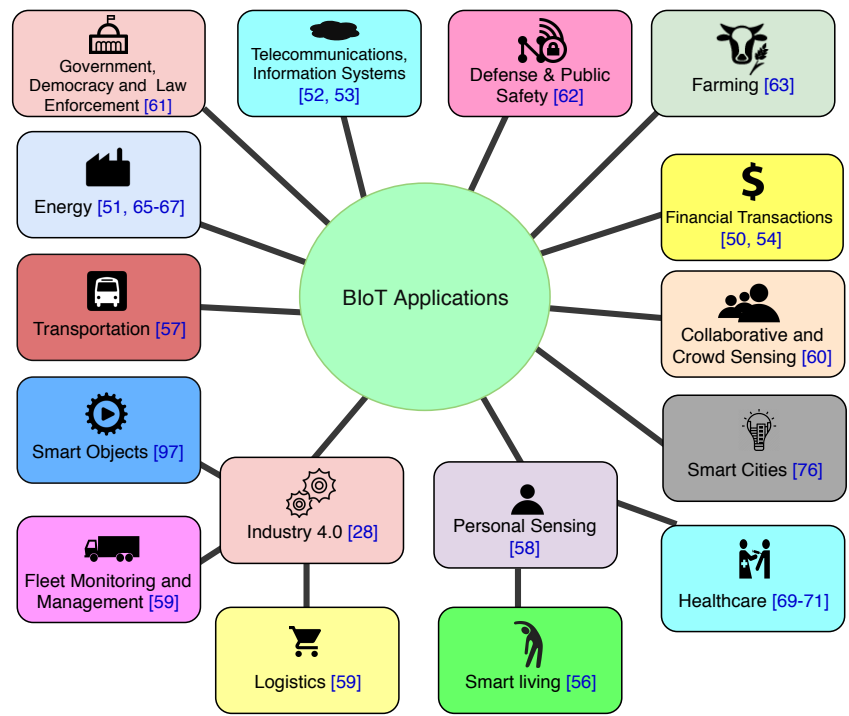

FIGURE 4: BIoT applications.

Other already proposed BIoT applications are related to smart cities [76] and industrial processes [28]. In the case of [76] it is proposed a framework that integrates smart devices in a secure way for providing smart city applications. In [28], different blockchain-based industrial applications are reviewed, including their connection to Industrial IoT (IIoT) networks.

Finally, it should be mentioned that Big Data can be leveraged by blockchain technology (i.e., to ensure its trustworthiness), so some researchers [27] reviewed the main blockchain-based solutions to gather and control massive amounts of data that may be collected from IoT networks.

## IV. DESIGN OF AN OPTIMIZED BLOCKCHAIN FOR IOT APPLICATIONS

Blockchain technologies can bring many benefits to IoT, but, since they have not been devised explicitly for IoT environments, the different pieces that make them up should be adapted. In order to optimize them, several authors studied BIoT performance in different scenarios. They analyzed a number of influential aspects, but they mainly focus on the performance of consensus algorithms.

An example of performance evaluation is detailed in [77]. Specifically, the paper analyzes whether the Practical Byzantine Fault Tolerance (PBFT) consensus algorithm (described later in Section IV-D) could be a bottleneck in networks with a large amount of peers. Actually, the tests described make use of up to 100 peers that interact with a blockchain based on IBM's Bluemix. The experiments measure the average time to reach a consensus and it can be observed how it grows as the number of peers increases.

The scalability of Proof-of-Work (PoW) and Byzantine Fault Tolerance (BFT) based consensus methods is compared in [78]. The author points out that, although Bitcoin has been a clear success, its poor scalability makes no sense today, since there are modern cryptocurrency platforms like Ethereum. In the paper it is suggested to improve PoW performance by mixing it with a BFT protocol. In addition, it is stated that the implementation of the consensus protocols in hardware is probably the most promising way for improving the performance of any consensus method.

Besides the consensus algorithm, other elements of the blockchain can be adapted to be used in IoT networks. Thus, in the next subsections the different parts of a blockchain are analyzed in order to determine possible optimizations.

### A. ARCHITECTURE
The architecture that supports a blockchain used for IoT applications should have to be adapted to the amount of traffic that such applications usually generate. This is a concern for traditional cloud-based architectures, which, as it is illustrated in Figure 5, evolved towards more complex edge and fog computing-based architectures. In such a Figure it can be observed that three architectures depend on a cloud, although, in practice, the dependency degree varies a great deal. In the case of a cloud-based architecture, the data collected by the Node Layer are forwarded directly to the cloud through IoT gateways without further processing that the one needed for protocol conversion (in case it is needed). There are also gateways that perform more sophisticated tasks (e.g., sensor fusion [79]), but in most cloud-centered applications, most processing is carried out in the cloud.

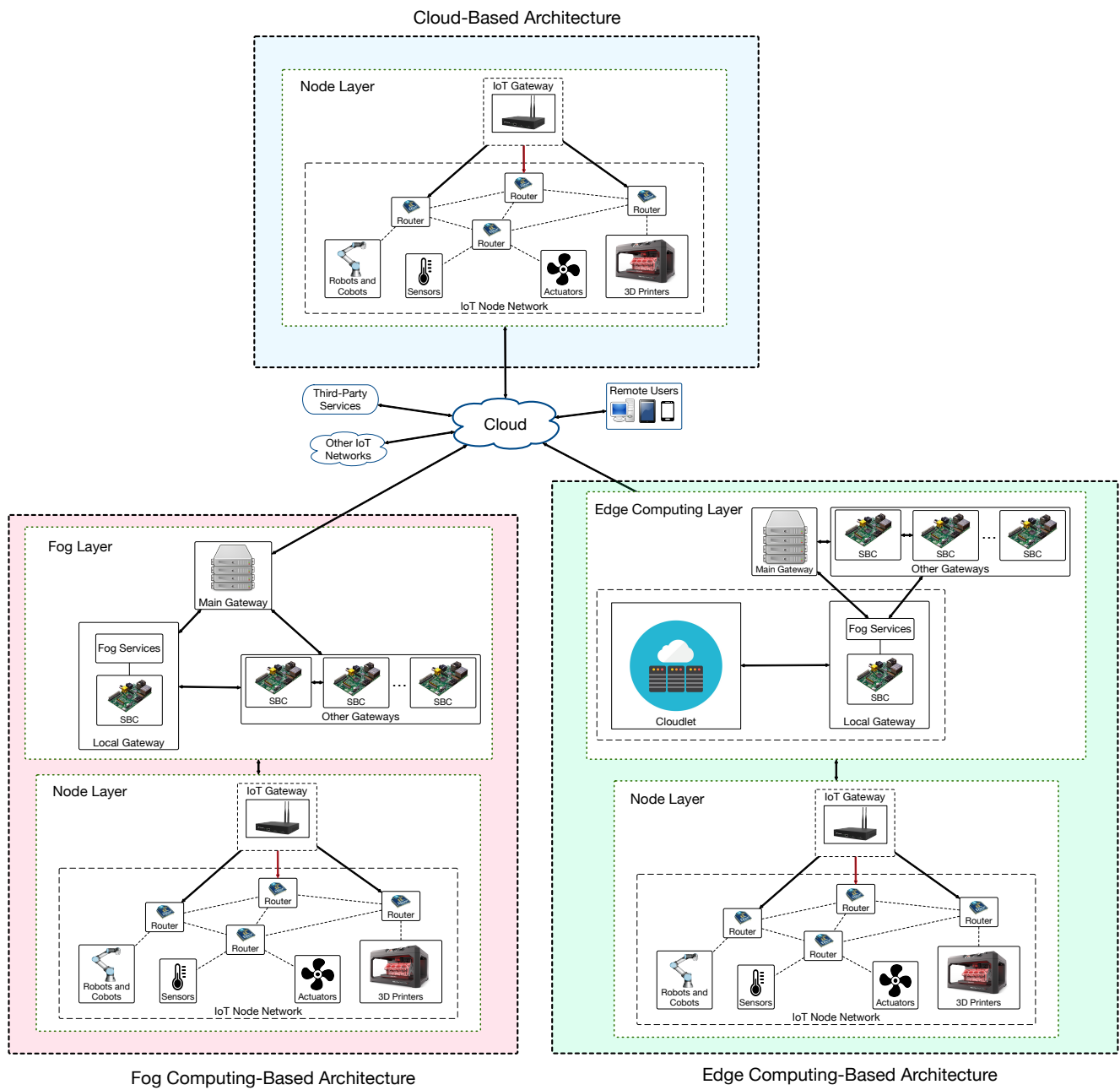

FIGURE 5: Traditional IoT architecture evolution.

However, note that traditional cloud-centered IoT architectures have certain inherent vulnerabilities [59], being the most relevant the fact that the cloud is a point of failure: if the cloud is down due to cyberattacks, maintenance or software problems, the whole system stops working. In addition, it is important to emphasize that if a single IoT device is compromised, it may disrupt the whole network by performing Denial of Service (DoS) attacks [80], eavesdropping private data [81], altering the collected data [82] or misleading other systems [83]. Therefore, once an IoT device connected to the cloud or to a central server is breached, the rest of the nodes may be compromised. In contrast, blockchain-based systems do not rely on a unique central server or cloud. Moreover, transactions are verified cryptographically, so when malicious activities from a compromised device are detected, the system can reject its blockchain updates.

The other two architectures depicted in Figure 5 are more recent and offload part of the processing from the cloud to the edge of the network. This offloading is key for IoT applications, since it is estimated that if the number of IoT connected devices keeps on growing at the same rate [1], the amount of communications to be handled by a cloud will increase remarkably and, therefore, the cloud network capacity will have to be expanded. Thus, Edge and fog computing can be used to support physically distributed, low-latency and QoS-aware applications that decrease the network traffic and the computational load of traditional cloud computing systems.

Fog computing is based on a set of local gateways able to respond fast to IoT node requests through specific services. Such nodes can also interact with each other and, when required, with the cloud (for instance, for long term storage). In Figure 5, fog local gateways are represented by Single-Board Computers (SBCs), which are low-cost and low-energy consumption computers that can be installed easily in a reduced space. Examples of popular SBCs are the different versions of Raspberry Pi [84] or BeagleBone [85].

Fog computing is actually considered a subset of edge computing [72], which has recently been presented as a valid architecture for supporting blockchain and blockchainless DAG IoT applications [86]. As it can be observed in Figure 5, in the Edge Computing Layer, besides fog gateways there is a cloudlet, which in practice consists in one or more high-end computers that act like a reduced version of a cloud. The main advantage of cloudlets is that they can provide high-speed responses to compute-intensive tasks required by the Node Layer (e.g., running a full node of a blockchain), which cannot be delivered effectively when using resource-constrained fog gateways.

There are other architectures that have been explored in the past in order to tackle the architectural issues that arise when providing BIoT services. A brief but good compilation of alternatives can be found in [87]. In such a paper the advantages and disadvantages of four different architectures (that the authors call Fully Centralized, Pseudo-Distributed Things, Distributed Things and Fully Distributed) are discussed. The researchers conclude that a BIoT architecture should be as close as possible to the Fully Distributed approach, but that, in some scenarios where computational power or cost are limiting factors, other approaches may be more appropriate.

An interesting platform that promotes decentralization for IoT systems is IBM's ADEPT. Such a platform was conceived for secure, scalable and autonomous peer-to-peer IoT telemetry. According to the authors, ADEPT is presented more as a starting point for discussion than as an implementation, but its white paper [88] provides a detailed description on the requirements for the platform. For instance, the researchers point out that an IoT device should be able to authenticate autonomously and to self-maintain, leaving to the manufacturers the responsibility of registering new devices in the blockchain. In addition, ADEPT's vision of mining is different from the one implemented in Bitcoin. Mining is necessary in Bitcoin to restrict currency issuance, but IBM considers that such a limitation restricts scalability and imposes an increasing computational cost. Therefore, ADEPT uses Proof-of-Stake (PoS) and PoW, which guarantee network integrity and security, but which do not impose additional limitations. Furthermore, it is worth mentioning that IBM's architecture for ADEPT distinguishes among three types of IoT devices (Light Peers, Standard Peers and Peer Exchanges), which differ in their role and computational capabilities. Finally, the authors of the white paper indicate the software selected for implementing ADEPT (Telehash [89], BitTorrent [90] and Ethereum) and describe different practical use cases of the system, like a washer that buys detergent automatically when it is low.

Another BIoT architecture is proposed in [91], [92]. In such papers the authors devise a theoretical lightweight architecture with security and privacy in mind, which reduces the communications overhead introduced by the use of a blockchain. The presented system is oriented towards home automation and its architecture is divided into three layers: the smart home layer, where there are sensors, actuators and local storage; an overlay network of peers and shared storage; and a cloud, which also provides remote storage. In the lower layers (smart homes and overlay network) storage is composed by traditional storage servers and blockchains, either public or private. The reduction in overhead is carried out by removing the PoW consensus mechanism, so every block is mined and appended to the blockchain without additional efforts. Every transaction is also appended to a block and is assumed that it is a true transaction, being the owner the one responsible for adding/removing devices. This simplification eases the blockchain functioning and, although the researchers studied the impact of different attacks on the system, it is not clear that the proposed scheme would withstand attacks performed by compromised IoT nodes whose contribution (e.g., collected sensor values), which is assumed to be true by default, may alter the behavior of other subsystems.

IoT is also gaining traction thanks to its global vision where devices are interconnected seamlessly among them and with the environment. For such a purpose, in [93] it is presented a theoretical blockchain-based architecture focused both on providing IoT services and connecting heterogeneous devices. The proposed architecture makes use of hierarchical and multi-layered blockchains, which enable building a contextual service discovery system called CONNECT.

A multi-layer IoT architecture based on blockchain technology is described in [94]. The proposed architecture decreases the complexity of deploying a blockchain by dividing the IoT ecosystem in levels and making use of the blockchain in each one. The researchers state that the architecture harnesses both the power of a cloud and the security and relia-bility of the blockchain.

A slightly different approach is presented in [95], where it is evaluated the use of a cloud and a fog computing architecture to provide BIoT applications. The authors indicate that the architecture is proposed because is really difficult to host a regular blockchain on traditional resource-constrained IoT devices. Thus, the researchers measure empirically the performance of the system proposed by using IoT nodes based on Intel Edison boards and IBM's Bluemix as blockchain technology. The obtained results show that, under high transaction loads, the fog system latency response is clearly faster than in a cloud-based system. Following similar ideas, the same authors presented another two works. In [96] they describe the implementation of RESTful microservices on the architecture, while in [97] they extend the architecture to a paradigm they call the Internet of Smart Things.

Another architecture based on edge computing is presented in [98], which describes ongoing research on the development of a hierarchical and distributed platform based on the IEC 61499 standard [99], which supports distributed automation control systems. Such systems can be structured in two layers: a bottom layer that controls devices and processes, and a top layer that supervises the bottom layer. For the top layer, the platform uses a blockchain based on Hyperledger Fabric [35] that implements smart contracts to perform supervision tasks. The edge nodes conform the bottom layer and are based on a micro-service architecture that makes use of Docker containers [100] and Kubernetes [101].

Software Defined Networking (SDN) has been also suggested for implementing BIoT architectures. For instance, in [102] it is proposed a novel blockchain-based architecture that makes use of SDN to control the fog nodes of an IoT network. The system makes use of a cloud to perform compute-intensive tasks, while providing low-latency data access through fog computing. The fog nodes are the ones that are distributed, providing services and interaction with the blockchain. The results obtained by the authors indicate that the architecture reduces delays, increases throughput and it is able to detect real-time attacks on the IoT network. In the specific case of a flooding attack, the architecture is able to balance the load between the fog nodes thanks to the use of the blockchain and an SDN algorithm. In addition, the same authors describe in [103] a similar SDN-based approach.

### B. CRYPTOGRAPHIC ALGORITHMS

Public-key cryptography is essential for providing security and privacy in a blockchain. However, resource-constraint IoT devices struggle with the computing requirements of modern secure cryptographic schemes [104]. Specifically, asymmetric cryptography based on Rivest–Shamir–Adleman (RSA) is slow and power consuming when implemented on IoT devices [41]. Therefore, when choosing the right cryptographic scheme, it should be taken into account not only the computational load and the memory requirements, but also the energy consumed.

The most common public-key based cipher suites are RSA and Elliptic Curve Diffie-Hellman Exchange (ECDHE), which are the ones recommended by the National Institute of Standards and Technology (NIST) [105] for Transport Layer Security (TLS) [106]. RSA-based cipher suites use RSA [107] as the key exchange algorithm, while the ECDHE-based ones use an algorithm that makes use of Ephemeral Diffie-Hellman based on Elliptic Curves [108].

Current RSA key sizes are not practical for most IoT devices. A 2048-bit key is the minimum size considered secure, since 768-bit and 1024-bit RSA implementations were broken in 2010 [109], [110]. Although possible, the use of a 2048-bit certificates on an ephemeral key exchange algorithm introduces heavy overhead and computing requirements, which are very difficult to accommodate on the constrained hardware capabilities of most IoT nodes.

In contrast, Elliptic Curve Cryptography (ECC) represents a much lighter alternative to RSA [111], [112]. It has already been shown that, when implemented on resource-constrained devices, ECC outperforms RSA in terms of speed [113]–[115] and power consumption [116]–[119]. However, note that in August 2015 the National Security Agency (NSA) recommended stopping the use of Suite B, an ECC-based algorithm, apparently, because of the progress recently made on quantum cryptography [120].

Regarding hash functions, they are also key in a blockchain-based system, since they are required to sign transactions. Therefore, hash functions for IoT applications have to be secure (i.e., they should not generate collisions [121]), fast and should consume the smallest possible amount of energy.

The most popular blockchain hash functions are SHA-256d (used by Bitcoin, PeerCoin or Namecoin), SHA-256 (used by Swiftcoin or Emercoin) and Scrypt (used by Litecoin, Gridcoin or Dogecoin). The performance of SHA-256 has been evaluated in different IoT devices, like wearables [122]. However, researchers that evaluated the footprint and energy requirements of SHA-256 in ASICs, concluded that, for low-power secure communications, it is more efficient to make use of Advanced Encryption Standard (AES) [123]. Due to such power limitations, other researchers suggested using ciphers like Simon [124], but further research and empirical evaluations on real BIoT applications are still needed.

### C. MESSAGE TIMESTAMPING

In order to track modifications on the blockchain, transactions have to be both signed and timestamped. This last task should be performed in a synchronized way, so timestamping servers are commonly used.

Different timestamping mechanisms can be used. Traditional schemes rely on having trustworthiness on the server, which signs and timestamps transactions with its own private key. Nonetheless, no one deters the server from signing past transactions. For such a reason, diverse authors have proposed secure mechanisms. For instance, the method implemented by Bitcoin is inspired by one of the solutions proposed in [125], where each timestamp includes a hash of the previous timestamp, what maintains the order of the transactions (even when the clocks are inaccurate) and makes it difficult to insert fake transactions in the already linked chain. In addition, timestamping can be distributed, hence avoiding the problem of having a single point of failure. Although such a distributed system is prone to Sybil attacks [126], Bitcoin solves them by linking blocks and using the PoW mechanism.

Other authors recently proposed the use of a decentralized timestamping service [127] or the distribution of its keys [128], but the topic has still to be studied in detail when decentralizing the service among devices of an IoT network.

### D. CONSENSUS MECHANISMS, MINING AND MESSAGE VALIDATION

Consensus is key for the proper functioning of a blockchain. It basically consists in a mechanism that determines the conditions to be reached in order to conclude that an agreement has been reached regarding the validations of the blocks to be added to the blockchain [26]. In practice, the problem is the Byzantine Generals Problem previously described in the Introduction.

The most egalitarian (and idealistic) consensus mechanism consists in giving to all the miners the same weight when voting and then deciding according to the majority of the votes. This scheme may be possible to implement in a controlled environment, but, in a public blockchain, this mechanism would lead to Sybil attacks, since a unique user with multiple identities would be able to control the blockchain [126].

In practice, in a decentralized architecture, one user has to be selected to add every block. This selection could be performed randomly, but the problem is that random selection is prone to attacks. PoW consensus algorithms are based on the fact that if a node performs a lot of work for the network, it is less likely that it is going to attack it. Specifically, the solution proposed by PoW-based blockchains makes it difficult to perform Sybil attacks by requiring miners to perform computationally expensive tasks that, theoretically, cannot be carried out by a single entity. The work performed usually involves doing some calculations until a solution is found, a process that is commonly known as mining. In the case of the Bitcoin blockchain, mining consists in finding a random number (called nonce) that will make the SHA-256 hash of the block header to have at the beginning certain number of zeroes. Therefore, miners have to demonstrate that they have performed certain amount of work to solve the problem. Once the problem is solved, it is really easy for other nodes to verify that the obtained answer is correct. However, this mining process makes the blockchain inefficient in throughput, scalability [78], and in terms of energy consumption, what is not desirable in an IoT network.

Due to the problems previously mentioned, several alternative consensus methods have been proposed. The following are the most relevant:

- PoS is a consensus mechanism that requires less computational power than PoW, so it consumes less energy. In a PoS-based blockchain it is assumed that the entities with more participation on the network are the ones less interested in attacking it. Thus, miners have to prove periodically that they own certain amount of participation on the network (e.g., currency). Since this scheme seems unfair, because the wealthiest participants are the ones ruling the blockchain, other variants have been proposed. For example, Peercoin's consensus algorithm [129] takes coin age into account: the entities with the oldest and largest sets of coins would be more likely to mine a block. Because of the advantages of PoS, some blockchains like Ethereum are planning to move from PoW to PoS.
- Delegated Proof-of-Stake (DPoS) [130] is similar to PoS, but stakeholders instead of being the ones generating and validating blocks, they select certain delegates to do it. Since less nodes are involved in block validation, transactions are performed faster than with other schemes. In addition, delegates can adjust block size and intervals, and, if they behave dishonestly, they can be substituted easily.
- Transactions as Proof-of-Stake (TaPoS) [131] is a PoS variant. While in PoS systems only some nodes contribute to the consensus, in TaPoS all nodes that generate transactions contribute to the security of the network.
- Proof-of-Activity (PoA) consensus algorithms were proposed due to the main limitation of PoS systems based on stake age: it is accumulated even when the node is not connected to the network. Thus, PoA schemes have been proposed to encourage both ownership and activity on the blockchain [132], rewarding stakeholders who participate instead of punishing passive stakeholders. A similar approach is proposed by Proof-of-Stake-Velocity (PoSV) [133]. It is implemented by Reddcoin [134], which is based on the concept of velocity of money. Such a concept indicates how many times a unit of currency flows through an economy and is used by the members of a society during a certain time period. Usually, the higher the velocity of money, the more transactions in which it is used and the healthier the economy.
- PBFT [135] is a consensus algorithm that solves the Byzantine Generals Problem for asynchronous environments. PBFT assumes that less than a third of the nodes are malicious. For every block to be added to the chain, a leader is selected to be in charge of ordering the transaction. Such a selection has to be supported by at least 2/3 of the all nodes, which have to be known by the network.
- Delegated BFT (DBFT) is a variant of BFT where, in a similar way to DPOS, some specific nodes are voted to be the ones generating and validating blocks.
- The Ripple consensus algorithm [136] was proposed to reduce the high latencies found in many blockchains,

which are in part due to the use of synchronous communications among the nodes. Thus, each Ripple's server (i.e., miner) relies on a trusted subset of nodes when determining consensus, what clearly reduces latency.
- Stellar Consensus Protocol (SCP) is a implementation of a consensus method called Federated Byzantine Agreement (FBA) [137]. It is similar to PBFT but, whilst in PBFT every node queries all the other nodes and waits for the majority to agree, in SCP the nodes only wait for a subset of the participants that they consider important.
- BFTRaft [138] is a BFT consensus scheme based on the Raft algorithm [139], which is aimed at being simple and easy to understand for students. Such an aim makes Raft assume simplifications that rarely hold in practice, like the fact that nodes only fail by stopping. Thus, BFTRaft enhances the Raft algorithm by making it Byzantine fault tolerant and by increasing its security against diverse threats.
- Sieve [140] is a consensus algorithm proposed by IBM Research that has already been implemented for Hyperledger-Fabric. Its objective is to run non-deterministic smart contracts on a permissioned blockchain that makes use of BFT replication. In such a scenario, Sieve replicates the processes related to non-deterministic smart contracts and then compares the results. If a divergence is detected among the results obtained by a small number of processes, they are sieved out. However, if the number of divergent processes is excessive, the whole operation is sieved out.
- Tendermint [141] is a consensus algorithm that can host arbitrary application states and can only tolerate up to a 1/3 of failures. In Tendermint, blockchain participants are called validators and they propose blocks of transactions and vote on them. A block is validated in two stages (pre-vote and pre-commit) and it can only be committed when more than 2/3 of the validators pre-commit it in a round.
- Bitcoin-NG [142] implements a variant of the Bitcoin consensus algorithm aimed at improving scalability, throughput and latency. The developers performed experiments with 1,000 nodes and concluded that Bitcoin-NG scales optimally, only limited by the bandwidth of the nodes and the latency related to the propagation time of the network.
- Proof-of-Burn (PoB) is a consensus method that requires miners to show proof of their commitment to mining by burning some cryptocurrency through an unspendable address. The idea behind PoB is that, instead of burning resources (e.g., energy in the case of many PoW implementations), cryptocurrency is burnt as it is considered as expensive as such resources.
- Proof-of-Personhood (PoP) [143] is a consensus mechanism that makes use of ring signatures [144] and collective signing [145] to bind physical to virtual identities in a way that anonymity is preserved. A very similar concept is Proof-of-Individuality (PoI), which is currently being developed on Ethereum by the PoI Project [146].

Finally, it is worth noting that private blockchains, which control user access, reduce the probability of Sybil attacks, so they do not require costly mining algorithms and economic incentives are removed.

### E. BLOCKCHAIN UPDATING/MAINTENANCE AND PROTOCOL STACK

The construction of an IoT network requires deploying a huge number of devices. Such devices embed certain firmware that is usually updated to correct bugs, prevent attacks [147] or just to improve some functionality. Traditionally, IoT devices had to be updated manually or with Over-The-Air (OTA) updates [148]. According to some researchers [149] these updates can be performed by using a blockchain, which enables IoT devices to spread securely new firmware versions.

Regarding the protocol stack, some authors suggested changes on the traditional OSI stack to adapt it to blockchain technologies. The most relevant is the so-called "Internet of Money" (IoM) [150], which proposes a set of five layers that operate on TCP/IP (shown in Figure 6). Such five layers include:
- A Ledger Layer that creates ledgers and issues assets.
- A Payment and Exchange Layer.
- A Pathfinding Layer that calculates the optimal set of atomic operations to be executed for the desired value transfer or exchange.
- A Contract Layer that controls balances through certain running code.

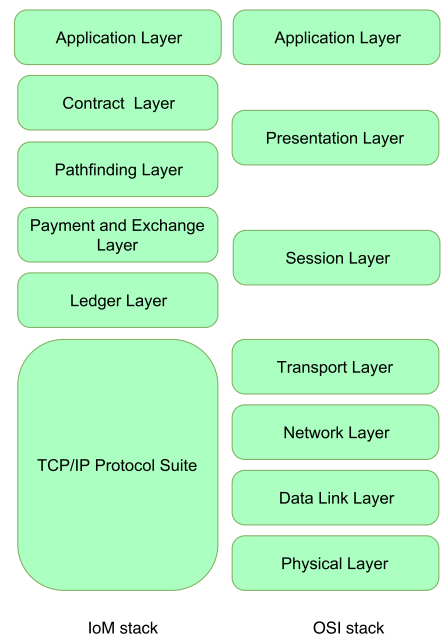

FIGURE 6: IoM versus traditional OSI protocol stack.

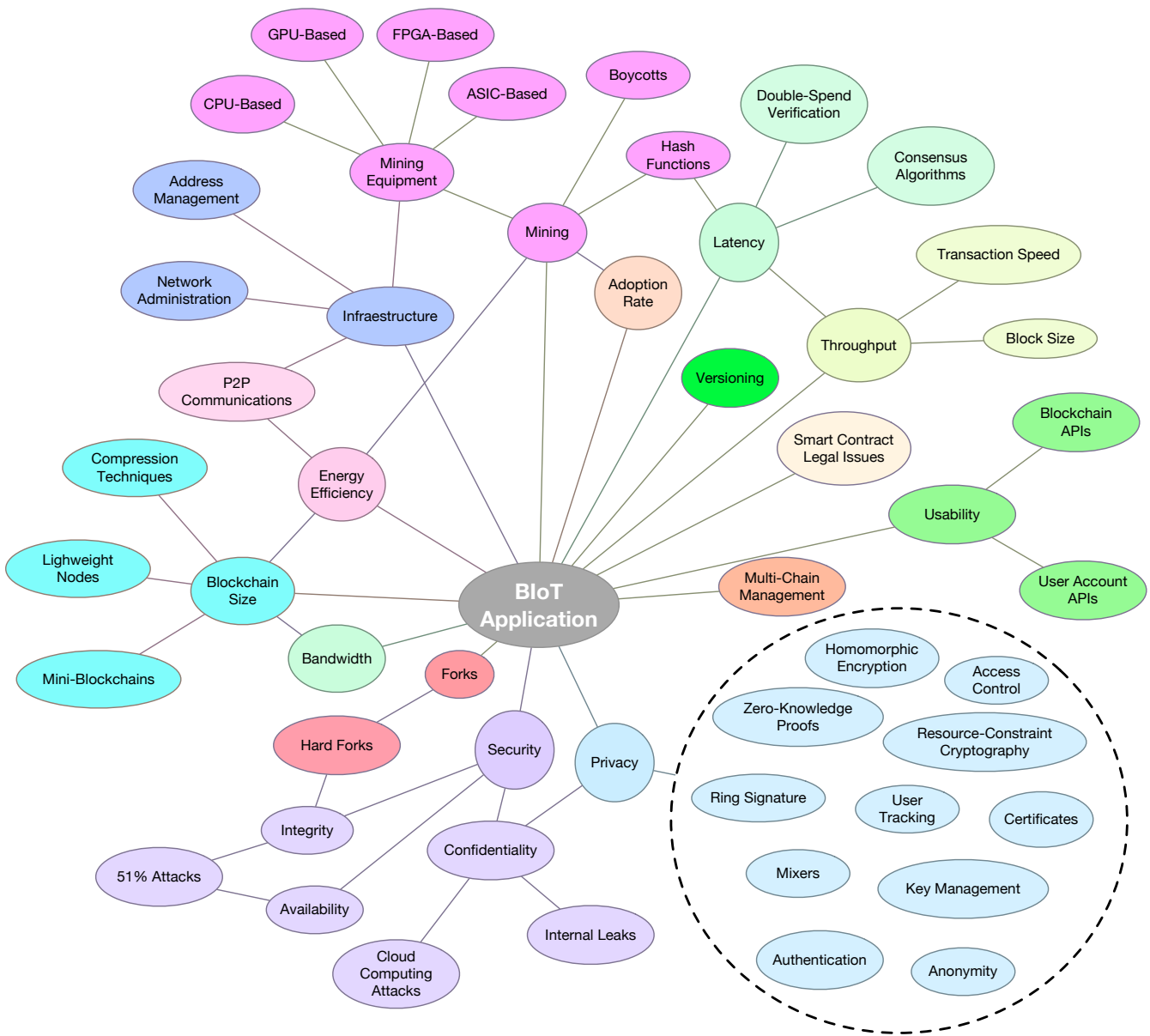

FIGURE 7: Most relevant factors that condition the development of a BIoT application and their main relationships.

- An Application Layer that allows for developing applications and user interfaces.

More research is still needed in order to study the need for specific stacks and to analyze their performance in comparison to other traditional OSI-based stacks.

## V. CURRENT CHALLENGES FOR BIOT APPLICATIONS

Today, emerging technologies in the IoT ecosystem like Cyber-Physical Systems (CPS) [151]–[153], RFID [154], telemetry systems [155] or 4G/5G broadband communications [156], [157] have to face several challenges. Specifically, the case of mission-critical scenarios [158] rise additional concerns. Adding blockchain to the mix implies further operational and technical requirements since the development of BIoT applications is a complex process that is affected by many aspects that are interrelated. The main factors are described in the next subsections and are depicted in Figure 7.

### A. PRIVACY

All the users of a blockchain are identified by their public key or its hash. This means that anonymity is not guaranteed and, since all transactions are shared, it is possible for third-parties to analyze such transactions and infer the actual identities of the participants [159], [160]. Privacy is even more complex in IoT environments, since IoT devices can reveal private user data that could be stored in a blockchain whose privacy requirements differ from one country to another [161].

Therefore, in contrast to traditional online payments, which are usually only visible to transacting parties and to a middleman (e.g., financial institutions, government), the transparent transactions fostered by blockchain are a challenge in terms of privacy.

Identity certification may also be a problem in IoT: if an identity provider is responsible for authorizing entities, it can also be able to block them. To address such a challenge, in [162] it is proposed the use of a permissioned blockchain for securing and managing multiple IoT nodes. The proposed system provides a distributed identity management solution that increases security and protection against attacks by rotating asymmetric keys. Such keys are generated locally on the device and they are never moved from it. To verify the identity of a user while rotating keys, the system makes use of a mechanism called Device Group Membership (DGM) that includes in a group all the devices that belong to a user and, when a user carries out a transaction, it is reflected on the blockchain as it was performed by a device that belonged to the user's group. The proposed solution also enhances security by using a certificate system for authentication and by enabling the hash function substitution if it is compromised. It is also worth mentioning that the system can be tweaked to limit the amount of temporal data stored, which is useful for IoT devices with little storage space (for instance, it could only be stored the data from the previous 24 hours).

Another approach focused on solving the privacy and robustness problems derived from using centralized identity management systems is described in [92]. There the authors emphasize the need for providing automatic authentication systems for IoT applications where scalability is needed and where device heterogeneity and mobility are common. To deal with such challenges, the researchers present a blockchain-based system for IoT smart homes that extracts appliance signatures automatically in order to identify both the appliances and their users.

Access management to IoT networks is challenging as well. Some researchers [163] suggested improving it by defining a blockchain-based multi-level mechanism, which would specify capabilities, access lists and access rights. However, note that, in many IoT applications anonymity is not necessary, but the privacy of the transactions is required in certain scenarios when the collected data may allow for monitoring and predicting people behavior or habits. This has already been an issue in fields like RFID-based transportation card systems, where the stored information (i.e., trips, balance, personal data) is supposedly anonymous, but in practice it may be collected by third parties [164]–[166]. The issue is even more problematic when adding a blockchain, since transactions are shared among peers, what in certain fields like industry or financial systems, allows for monitoring the activity of competitors.

Therefore, solutions have to be proposed to mitigate these privacy issues. For example, in the case of public blockchains a user does not need to know the address of every user, just the one of the counterparty he/she is dealing with. If a blockchain participant makes use of a new address for every transaction, data analysis would become more difficult. This is similar to what smartphones manufacturers have implemented to avoid Wi-Fi tracking [167], [168]. A more practical but less anonymous solution would consist in using a unique address for each counterparty.

In a private blockchain, since access controls are performed, there is at least one node that knows who accesses the system. Assuming the neutrality of the access controller, it is possible to reduce exposure by establishing an independent blockchain with every entity a user is collaborating with. This setup increases communications complexity, but isolates the user from non-desired monitoring. For instance, Multichain [169] provides a solution for deploying private blockchains (it can work with different blockchains at the same time) that ensures that the activities on the blockchain can be monitored by chosen participants.

Mixing techniques can also help to enhance privacy. Such techniques can collect transactions from diverse IoT devices and output events or other transactions to different addresses that are not linked to the original devices. These techniques increase privacy, but they are not perfect, since they may be de-anonymized through statistical disclosure attacks [170]. Moreover, the mixing service has to be trusted, since a malicious mixer may expose users and, in the case of economic transactions, it may end up stealing coins. To tackle these issues different proposals suggested exposing theft through an accountability mechanism [171] or hiding the input/output address mapping from the mixing server [172].

Privacy can also be increased through zero-knowledge proving techniques like the ones used by Zerocoin [173], Zerocash [174] or Zcash [175]. A zero-knowledge proof is a method that allows for proving to a counterparty that a user knows certain information without revealing such an information [176]. In the case of IoT applications, zero-knowledge proofs can be used for authentication or during regular transactions in order to avoid revealing the identity of a user or a device. However, note that these proofs are not immune to attacks [177]. In fact, like in the case of mixing techniques, they are susceptible to de-anonymization through statistical disclosure attacks, but they improve mixing techniques by avoiding the necessity for a mixing server, which can pose a security or performance bottleneck.

It must be also remarked the privacy-focused efforts performed by several initiatives like Bytecoin [178] or Monero [179], which are based in CryptoNote [180]. CryptoNote is a protocol that makes use of ring signatures and whose transactions cannot be followed through the blockchain in order to determine who performed them. The only people that can access the transaction information are the parties that carry it out or whoever knows one of the two private keys. One of the keys of CryptoNote is its implementation of the concept of ring signature [144], which makes it possible to specify a set of possible signers without revealing who of them actually produced the signature.

Another possible solution for preserving privacy is the use of homomorphic encryption [181], [182]. Such a kind of encryption allows third-party IoT services to process a transaction without revealing the unencrypted data to those services. Several researchers have suggested variations on the Bitcoin protocol to make use of homomorphic commitments [183], [184].

Finally, note that part of the mechanisms previously mentioned require a relevant number of computational resources, so its applicability to resource-constrained IoT devices is currently limited.

### B. SECURITY

Traditionally, three requirements have to be fulfilled by an information system in order to guarantee its security:

- Confidentiality. The most sensitive information should be protected from unauthorized accesses.
- Integrity. It guarantees that data are not altered or deleted by unauthorized parties. It is also usually added that, if an authorized party damages the information, it should be possible to undo the changes.
- Availability. Data can be accessed when needed.

Regarding confidentiality, the part related to the transaction data is associated with their privacy, which has been already analyzed in the previous subsection. With respect to the infrastructure that supports the stored data, it can be stated that current IoT applications tend to centralize communications in a server, in a farm of servers or in a cloud. Such an approach is valid as long as the administrators of the centralized infrastructure are trusted and while the system remains robust against attacks [185], [186] and internal leaks. In contrast, blockchain technologies are characterized by being decentralized, so, although one node is compromised, the global system should keep on working.

For an individual user, the key for maintaining confidentiality is a good management of his/her private keys, since it is what an attacker needs in conjunction with the public key to impersonate someone or steal something from him/her. An interesting initiative related to this topic is CONIKS [187], a key management system created to liberate users from encryption key management. In such a system the user first has to ask for a public key to a provider, which only requires a user name to register in the CONIKS system. When a user wants to send a message to another user, his/her CONIKS client looks for the counterparty's key in the key directory. In order to avoid key tampering from the service provider (which might become compromised), before sending any message, two verifications are performed: it is checked that the public key of the receiver is the one used by other clients when communicating with the same user, and that such a key has not changed unexpectedly over time. Similar solutions have been proposed for IoT devices, making use of blockchain technology to strengthen their identity and access management, since blockchains provide a defense against IP spoofing and forgery attacks [59].

Certificates are also essential when guaranteeing security on the Internet. Therefore, certificate authorities that make use of a public-key infrastructure have to provide trust to third-parties. However, such authorities have proven to fail in certain occasions [188], then having to invalidate certificates previously issued. Some recent initiatives are aimed at fixing certain structural flaws found in the SSL certificate system. Specifically, Google's Certificate Transparency [189] provides a framework for monitoring and auditing SSL certificates in almost real time. The solution uses a distributed system based on Merkle hash trees that allows third-parties to audit and verify whether a certificate is valid.

With respect to integrity, it must be indicated that the foundations of a blockchain are designed to store information that cannot be altered (or that it is very costly to do it) once it is stored. Nonetheless, note that in the past there were certain situations when this principle was ignored. For instance, in 2014, in an event that it is still to be clarified, the currency exchange platform MintPal notified its users that a hacker had stolen almost 8 million Vericoins, what was about 30% of the total coins of such a platform. To prevent the loss of investor funds and the fact that an actor would control 30% of the coin's proof-of-stake network capacity, the Vericoin developers decided to hard fork the blockchain, reversing the damage (a hard fork is a permanent divergence from the previous version of the blockchain). Therefore, although many information sources indicate that blockchains are a permanent storage for data that cannot be altered, it is actually not true in practice for preserving integrity in very exceptional cases. In IoT applications, data integrity is also essential and it is usually provided by third-parties. To avoid such a dependence, in [190] it is proposed a data integrity service framework for cloud-based IoT applications that makes use of blockchain technology, thus eliminating the need for trusting such third-parties.

The third characteristic of security is availability, but it is actually the most straightforward to be fulfilled by blockchains, since they are conceived by design to be distributed systems, what allows them to keep on working even when some nodes are under attack. Nevertheless, availability can be compromised through other types of attacks. The most feared attack is a 51-percent attack (also called majority attack), where a single miner can control the whole blockchain and perform transactions at wish. In this situation, data are available, but the availability for performing transactions can be blocked by the attacker that controls the blockchain. Obviously, this kind of attack also affects data integrity.

### C. ENERGY EFFICIENCY

IoT end-nodes usually make use of resource-constrained hardware that is powered by batteries. Therefore, energy efficiency is key to enable a long-lasting node deployment. However, many blockchains are characterized by being power-hungry. In such cases most of the consumption is due to two factors:

- Mining. Blockchains like Bitcoin make use of massive amounts of electricity due to the mining process, which

- involves a consensus algorithm (PoW) that consists in a sort of brute force search for a hash.
- P2P communications. P2P communications require edge devices that have to be powered on continuously, which could lead to waste energy [191], [192]. Some researchers proposed energy efficient protocols for P2P networks [193]–[195], but the issue still has to be studied further for the specific case of IoT networks.

Regarding mining, some authors suggested that the power consumed by proofs of work could be used for something useful while providing at the same time the required PoW [196]. Obtaining such proofs should have certain degree of difficulty, while its verification should be really fast. Some initiatives based on blockchains, like Gridcoin [197], reward volunteer scientific research computing with coins (although, as a consensus algorithm, Gridcoin uses PoS). Another interesting example is Primecoin [198], whose PoW mechanism looks for chains of prime numbers. Thus, a massive infrastructure like the one involved in IoT could also be harnessed to solve problems while making use of a blockchain.

Proof-of-Space (PoS) (also known as Proof-of-Capacity (PoC)) has also been suggested as a greener alternative to PoW [199]. PoS systems require users to show a legitimate interest in a specific service by allocating certain amount of memory or disk. This mechanism has already been implemented by cryptocurrencies like Burst-coin [200]. Other consensus methods have been proposed to reduce energy consumption respect to PoW, like Proof-of-Stake or Practical Byzantine Fault Tolerance (both described in Section IV-D).

In relation to P2P communications, they are essential for a blockchain to communicate peers and distribute blocks, so the more updates a blockchain receives, the more energy consumption is dedicated to communications. To reduce the number of updates, mini-blokchains [201] may allow IoT nodes to interact directly with a blockchain, since they only keep the latest transactions and lower the computational requirements of a full node.

In terms of hashing algorithms, SHA-256 is the reference due to being the one used by Bitcoin, but new algorithms like Scrypt [202] or X11 [203] are faster and thus can reduce mining energy consumption. Other hashing algorithms have been suggested, like Blake-256 [204], and some blockchains are able to make use of different hashing algorithms (e.g., Myriad [205]), but further analyses should be carried out on the performance and optimization of modern hash functions to be used on IoT devices.

### D. THROUGHPUT AND LATENCY

IoT deployments may require a blockchain network able to manage large amounts of transactions per time unit. This is a limitation in certain networks. For instance, Bitcoin's blockchain has a theoretical maximum of 7 transactions per second [78], although it can be increased by processing larger blocks or by modifying certain aspects of the node behavior when accepting transactions [206]. In comparison, other networks are remarkably faster. For instance, VISA network (VisaNet) can handle up to 24,000 transactions per second [207].

Regarding latency, it is important to note that blockchain transactions take some time to be processed. For example, in the case of Bitcoin, block creation times follow a Poisson distribution with a 10-minute mean [18], although, for avoiding double-spend, merchants are recommended to wait for about an hour, since five or six blocks usually need to be added to the chain before the transaction is confirmed. This latency requires only a few seconds in the case of VISA [207].

In relation to the consensus latency, it can be stated that the complexity of the consensus process is more important in terms of latency than individual hashing, but different blockchains, like the one that supports Litecoin [34], have opted for using scrypt, a hashing algorithm that is slightly faster than SHA-256.

### E. BLOCKCHAIN SIZE, BANDWIDTH AND INFRASTRUCTURE

Blockchains grow periodically as users store their transactions, what derives into larger initial download times and in having to make use of more powerful miners with larger persistent memories. Blockchain compression techniques should be further studied, but the truth is that most IoT nodes would not be able to handle even a small fraction of a traditional blockchain. Moreover, note that many nodes have to store large amounts of data that are of no interest for them, what can be regarded as a waste of computational resources. This issue could be avoided by using lightweight nodes, which are able to perform transactions on the blockchain, but who do not have to store it. However, this approach requires the existence in the IoT hierarchy of certain powerful nodes that would maintain the blockchain for the resource-constrained nodes, what implies a certain degree of data centralization.

Another alternative would consist in the use of a mini-blockchain [183], [201]. Such a kind of blockchain introduces the use of an account tree, which stores the current state of every user of the blockchain. Thus, only the most recent transaction has to be stored on the blockchain together with the account tree. Therefore, the blockchain only grows when new users are added to the blockchain.

In addition, note that transaction and block size have to be scaled according to the bandwidth limitations of IoT networks: many small transactions would increase the energy consumption associated with communications, while a few large ones may involve big payloads that cannot be handled by some IoT devices.

Regarding the infrastructure, certain elements are required to make the blockchain work properly, including decentralized storage, communication protocols, mining hardware, address management or network administration. Part of these needs are being fulfilled by the industry progressively, creating specific equipment for blockchain applications. For instance, miners have evolved from simple CPU-based systems, to more sophisticated equipment that harnesses the power of Graphics Processing Units (GPUs),

Field-Programmable Gate Arrays (FPGAs) or Application-Specific Integrated Circuits (ASICs) [208].

### F. OTHER RELEVANT ISSUES

#### 1) Adoption rate

One of the factors that may prevent a wide adoption of a BIoT application is the fact that a blockchain enables pseudo-anonymity (i.e., users or devices are identified by addresses, but they are not clearly linked to them). Governments may demand a strong link between real-world and online identity. Moreover, since IoT transactions can be carried out internationally, it may not be clear who should perform the identification.

In addition, note that the value and security of a blockchain increases with the number of users, also being more difficult to perform the feared 51-percent attacks.

Moreover, note that miner adoption rate also influences the capacity of a network to process transactions, so, in a BIoT deployment, the computational power brought by miners should be high enough to handle the transactions received from the IoT devices.

#### 2) Usability

In order to ease the work of developers, blockchain access Application Programming Interface (APIs) should be as user-friendly as possible. The same should be applied to the APIs to manage user accounts.

#### 3) Multi-chain management

In some cases, the proliferation of blockchains has derived into the necessity of having to deal with several of them at the same time. This can also happen in an IoT scenario, where, for instance, sensor values may be stored in a private blockchain, while financial transactions among nodes that provide services may be supported by Ethereum's or Bitcoin's blockchain.

#### 4) Versioning and forks

Blockchains can be forked for administrative or versioning purposes. Once a blockchain is forked, it is not easy to carry out transactions between both chains.

#### 5) Mining boycott

Miners end up deciding which transactions are or are not stored in the blockchain, so they are able to censor certain transactions for economic or ideological reasons. This issue can happen when the number of conspiring miners are above 51 percent of the total, so small chains and blockchains that delegate their decisions on a subset of miners are susceptible to this kind of boycotts. Therefore, miners have to be chosen wisely and, when smart contracts have been signed, misbehaviors should be sanctioned.

#### 6) Smart contract enforcement and autonomy

Legal rules have still to be developed to enforce smart contracts and resolve disputes properly. Some work is being performed for binding real-world contracts with smart contracts [161], but this is still an issue to be further studied.

## VI. FURTHER CHALLENGES AND RECOMMENDATIONS

Despite the promising benefits and the brilliant foreseen future of BIoT, there are significant challenges in the development and deployment of existing and planned systems that will need further investigation:

- Complex technical challenges: there are still issues to be addressed regarding the scalability, security, cryptographic development and stability requirements of novel BIoT applications. Moreover, blockchain technologies face design limitations in transaction capacity, in validation protocols or in the implementation of smart contracts. Furthermore, methods to solve the tendency to centralized approaches should be introduced.
- Interoperability and standardization: the adoption of BIoT will require the compromise of all stakeholders in order to achieve full interoperability (i.e., from data to policy interoperability) and integration with legacy systems. The adoption of collaborative implementations and the use of international standards for collaborative trust and information protection (i.e., access control, authentication and authorization) will be needed. For instance, authentication across multiple authorities or organizations requires Federated Identity Management (FIM) [209]. At an international scale, such a FIM currently exists only at a low Level of Assurance (LoA). The required LoA (from LoA 1 to LoA 4), as defined by the ISO/IEC 29115:2013 standard, is mainly based on risks, on the consequences of an authentication error and/or the misuse of credentials, on the resultant impact, and on their likelihood of occurrence. Thus, higher LoAs will be needed.
- Blockchain infrastructure: it will be needed to create a comprehensive trust framework or infrastructure that can fulfill all the requirements for the use of blockchain in IoT systems. Many state-of-the-art approaches that address issues such as trust depend on inter-domain policies and control. For instance, the governments should set up a blockchain infrastructure to support use cases of public interest.
- Organizational, governance, regulatory and legal aspects: besides technological challenges, shaping the regulatory environment (i.e., decentralized ownership, international jurisdiction) is one the biggest issues to unlock the potential value of BIoT. For instance, it is possible that some developers fake their blockchain performance in order to attract investors driven by the expected profits.
- Rapid field testing: in the near future, different types of blockchains for diverse applications will need to be optimized. Moreover, when users want to combine blockchain with IoT systems, the first step is to figure out which blockchain fits their requirements. Therefore, it is necessary to establish a mechanism to test dif-ferent blockchains. This approach should be split into two main phases: standardization and testing. In the standardization phase, after a wide understanding of the supply chains, markets, products, and services, all the requirements have to be analyzed and agreed. When a blockchain is created, it should be tested with the agreed criteria to verify if the blockchain works as needed. In the case of the testing phase, different criteria should be evaluated in terms of privacy, security, energy efficiency, throughput, latency, blockchain capacity or usability, among others.

## VII. CONCLUSIONS

The transition to a data-driven world is being accelerated by the pace of the technological advances of an Internet-enabled global world, the rise of societal challenges, and an increasing competition for scarce resources. In this ecosystem, blockchain can offer to IoT a platform for distributing trusted information that defy non-collaborative organizational structures.

This review examined the state-of-the art of blockchain technologies and proposed significant scenarios for BIoT applications in fields like healthcare, logistics, smart cities or energy management. These BIoT scenarios face specific technical requirements that differ from implementations involving cryptocurrencies in several aspects like energy efficiency in resource-constrained devices or the need for a specific architecture.

The aim of this work was to evaluate the practical limitations and identify areas for further research. Moreover, it presented a holistic approach to BIoT scenarios with a thorough study of the most relevant aspects involved in an optimized BIoT design, like its architecture, the required cryptographic algorithms or the consensus mechanisms. Furthermore, some recommendations were provided with the objective of giving some guidance to future BIoT researchers and developers on some of the issues that will have to be tackled before deploying the next generation of BIoT applications.

We can conclude that, as in any technological innovation, there is no one-size-fits-all solution for a BIoT application. Nevertheless, the adoption of the paradigm opens a wide area of short- and medium-term potential applications that could disrupt the industry and probably, the economy, as we know it today. The global reality is a complex mix of different stakeholders in the IoT ecosystem, therefore it is necessary to reassess the different activities and actors involved in the near-future economy. We can conclude that BIoT is still in its nascent stage, and beyond the earliest BIoT developments and deployments, broader use will require additional technological research advances to address the specific demands, together with the collaboration of organizations and governments.

## REFERENCES

[1] Gartner. Report: "Forecast: The Internet of Things, Worldwide, 2013". Nov. 2013.
[2] Cisco Systems. White paper: Cisco Visual Networking Index: Global Mobile Data Traffic Forecast Update, 2016–2021. March 2017.
[3] Suárez-Albela, M., Fraga-Lamas, P., Fernández-Caramés, T.M., Dapena, A., González-López, M. "Home Automation System Based on Intelligent Transducer Enablers", in Sensors, vol. 16 (10), no. 1595, pp. 1–26, September 2016.
[4] Fraga-Lamas, P., Fernández-Caramés, T. M., Castedo, L. "Towards the Internet of Smart Trains: A Review on Industrial IoT-Connected Railways" in Sensors, vol. 17 (6), no. 1457, pp.1–44, June 2017.
[5] Fraga-Lamas, P., Fernández-Caramés, T.M., Suárez-Albela, M., Castedo, L., González-López, M. "A Review on Internet of Things for Defense and Public Safety", in Sensors, vol. 16 (10), no. 1644, pp. 1–44, October 2016.
[6] Barro-Torres, S.J., Fernández-Caramés, T.M., Pérez-Iglesias, H.J., Escudero, C.J. "Real-Time Personal Protective Equipment Monitoring System", in Comput. Commun., vol. 36, pp. 42–50, 2012.
[7] Blanco-Novoa, Ó., Fernández-Caramés, T. M., Fraga-Lamas, P., Vilar-Montesinos, M. A. "A Practical Evaluation of Commercial Industrial Augmented Reality Systems in an Industry 4.0 Shipyard," in IEEE Access, vol. 6, pp. 8201-8218, 2018.
[8] Fraga-Lamas, P., Fernández-Caramés, T. M., Blanco-Novoa, Ó., Vilar-Montesinos, M. A. "A Review on Industrial Augmented Reality Systems for the Industry 4.0 Shipyard," in IEEE Access, vol. 6, pp. 13358-13375, 2018.
[9] Triantafillou, P., Ntarmos, N., Nikoletseas, S., Spirakis, P. "NanoPeer networks and P2P worlds", in Proceedings of the Third International Conference on Peer-to-Peer Computing, Linkoping, Sweden, 1-3 Sep. 2003.
[10] Ali, M., Uzmi, Z. A. "CSN: a network protocol for serving dynamic queries in large-scale wireless sensor networks", in Proceedings of the Second Annual Conference on Communication Networks and Services Research, Fredericton, Canada, 21 May 2004.
[11] Krco, S., Cleary, D., Parker, D. "P2P Mobile Sensor Networks", in Proceedings of the 38th Annual Hawaii International Conference on System Sciences, Hawaii, United States, 6 Jan. 2005.
[12] IBM. Executive report: "Device democracy: Saving the future of the Internet of Things". 2015.
[13] Landau, S."Making Sense from Snowden: What's Significant in the NSA Surveillance Revelations", in IEEE Security and Privacy, vol. 11, no. 4, pp. 54-63, July-Aug. 2013.
[14] Landau, S. "Highlights from Making Sense of Snowden, Part II: What's Significant in the NSA Revelations", in IEEE Security and Privacy, vol. 12, no. 1, pp. 62-64, Jan.-Feb. 2014.
[15] MarketsandMarkets; Statista estimates. Market for blockchain technology worldwide. Available online: https://www.statista.com/statistics/647231/worldwide-blockchain-technology-market-size/ (Accessed on 10 April 2018).
[16] Blockchain Technology report to the US Federal Advisory Committee on Insurance. Available online: https://www.treasury.gov/initiatives/fio/Documents/McKinsey_FACI_Blockchain_in_Insurance.pdf (Accessed on 10 April 2018).
[17] Crypto-currency market capitalizations. Available online: https://coinmarketcap.comhttps://coinmarketcap.com/ (Accessed on 10 April 2018).
[18] Nakamoto, S. "Bitcoin: A Peer-to-Peer Electronic Cash System". Available online: https://bitcoin.org/bitcoin.pdf (Accessed on 10 April 2018)
[19] Swan, M. "Blockchain: blueprint for a new economy". First Edition, O'Reilly Media, Jan. 2015.
[20] Singh, S., Singh, N. "Blockchain: Future of financial and cyber security", in Proceedings of the 2nd International Conference on Contemporary Computing and Informatics (IC3I), Noida, India, 14-17 Dec. 2016, pp. 463-467.
[21] Tschorsch, F., Scheuermann, B. "Bitcoin and Beyond: A Technical Survey on Decentralized Digital Currencies", in IEEE Communications Surveys & Tutorials, vol. 18, no. 3, Mar. 2016, pp. 2084-2123.
[22] Ethereum official web page. Available online: https://www.ethereum.org (Accessed on 10 April 2018)
[23] Counterparty official web page. Available online: wwww.counterparty.io (Accessed on 10 April 2018)
[24] Lamport, L., R. Shostack, and M. Pease. "The Byzantine Generals Problem", ACM Transactions on Programming Languages and Systems, vol. 4, no. 3, pp. 382–401, 1982.

[25] Christidis, K., Devetsikiotis, M., "Blockchains and Smart Contracts for the Internet of Things", in IEEE Access, vol. 4, pp. 2292-2303, May 2016.

[26] Zheng, Z., Xie, S., Dai, H., Chen, X., Wang, H. "An Overview of Blockchain Technology: Architecture, Consensus, and Future Trends", in Proceedings of the IEEE International Congress on Big Data (BigData Congress), Honolulu, United States, 25-30 June 2017, pp. 557-564.

[27] Karafiloski, E., Mishev, A., "Blockchain solutions for big data challenges: A literature review", in Proceedings of the IEEE International Conference on Smart Technologies, Ohrid, Macedonia, 6-8 July 2017.

[28] Ahram, T., Sargolzaei, A., Sargolzaei, S., Daniels, J., Amaba, B., "Blockchain technology innovations", in Proceedings of the IEEE Technology & Engineering Management Conference (TEMSCON), Santa Clara, United States, 8-10 June 2017.

[29] Conoscenti, M., Vetrò, A., De Martin, J. C., "Blockchain for the Internet of Things: A systematic literature review", in Proceedings of the IEEE/ACS 13th International Conference of Computer Systems and Applications (AICCSA), Agadir, Morocco, 29 Nov. - 2 Dec. 2016.

[30] Yli-Huumo, J., Ko, D., Choi, S., Park, S., Smolander, K. "Where Is Current Research on Blockchain Technology? - A Systematic Review", in PLOS ONE vol.11, no. 10, pp.1-27, 2016.

[31] Raval, S. Decentralized Applications: harnessing Bitcoin's blockchain technology. First edition, O'reilly Media, Aug. 2016.

[32] Mel, H.X., Baker, D. "Cryptography Decrypted", Addison Wesley, ISBN 0-201-61647-5, 2001.

[33] Ferguson, N., Schneier, B. "Practical Cryptography", Wiley, ISBN 0-471-22357-3, 2003.

[34] Litecoin official web page. Available online: https://litecoin.com (Accessed on 10 April 2018)

[35] Hyperledger-Fabric official web page. Available online: https://www.hyperledger.org/projects/fabric (Accessed on 10 April 2018)

[36] Ripple's official web page. Available online: https://www.ripple.com (Accessed on 10 April 2018)

[37] IOTA's official web page. Available online: https://www.iota.org (Accessed on 10 April 2018)

[38] Gui, T., Ma, C., Wang, F., Wilkins, D. E. "Survey on swarm intelligence based routing protocols for wireless sensor networks: An extensive study", in Proceedings of the IEEE International Conference on Industrial Technology (ICIT), Taipei, Taiwan, 14-17 March 2016.

[39] Preden, J. S., Tammemäe, K., Jantsch, A., Leier, M., Riid, A., Calis, E. "The Benefits of Self-Awareness and Attention in Fog and Mist Computing", in Computer, vol. 48, no. 7, pp. 37-45, July 2015.

[40] Bonomi, F., Milito, R., Zhu, J., Addepalli, S. "Fog Computing and its Role in the Internet of Things", in Proceedings of the First Edition of the MCC Workshop on Mobile Cloud Computing, Helsinki, Finlad, 17 Aug. 2012, pp. 13-16.

[41] Suárez-Albela, M., Fernández-Caramés, T. M., Fraga-Lamas, P., Castedo, L. "A Practical Evaluation of a High-Security Energy-Efficient Gateway for IoT Fog Computing Applications", in Sensors, vol. 17 (9), no. 1978, pp.1–39, August 2017.

[42] Datla, D., Chen, X., Tsou, T., Raghunandan, S., Shajedul Hasan, S. M., Reed, J. H., Dietrich, C. B., Bose, T., Fette, B., Kim, J.-H. "Wireless distributed computing: a survey of research challenges", in IEEE Communications Magazine, vol. 50, no. 1, pp. 144-152, Jan. 2012.

[43] Wu, Z., Meng, Z., Gray, J. "IoT-Based Techniques for Online M2M-Interactive Itemized Data Registration and Offline Information Traceability in a Digital Manufacturing System", in IEEE Transactions on Industrial Informatics, vol. 13, no. 5, pp. 2397-2405, Oct. 2017.

[44] Lomotey, R. K., Pry, J., Sriramoju, S., Kaku, E., Deters, R. "Wearable IoT Data Architecture", in Proceedings of theIEEE World Congress on Services (SERVICES), Honolulu, United States, 25-30 June 2017.

[45] Cai, H., Xu, B., Jiang, L., Vasilakos, A. V. "IoT-Based Big Data Storage Systems in Cloud Computing: Perspectives and Challenges", in IEEE Internet of Things Journal, vol. 4, no. 1, pp. 75-87, Feb. 2017.

[46] Marjani, M., Nasaruddin, F., Gani, A., Karim, A., Hashem, I. A. T., Siddiqa, S., Yaqoob, I. "Big IoT Data Analytics: Architecture, Opportunities, and Open Research Challenges", in IEEE Access, vol. 5, pp. 5247-5261, Mar. 2017.

[47] Back, A., Corallo, M., Dashjr, L., Friedenbach, M., Maxwell, G., Miller, A., Poelstra, A., Timón, J., Wuille, P. "Enabling Blockchain Innovations with Pegged Sidechains". Available online: https://www.blockstream.com/sidechains.pdf (Accessed on 10 April 2018)

[48] Pérez-Expósito, J., Fernández-Caramés, T. M., Fraga-Lamas, P., Castedo, L. "VineSens: An Eco-Smart Decision-Support Viticulture System", in Sensors, vol. 17 (3), no. 465, pp. 1–26, February 2017.

[49] Swanson, T., "Consensus-as-a-service: a brief report on the emergence of permissioned, distributed ledger system". Available online: http://www.ofnumbers.com/wp-content/uploads/2015/04/Permissioned-distributed-ledgers.pdf (Accessed on 10 April 2018)

[50] Wörner, D., von Bomhard, T., "When your sensor earns money: exchanging data for cash with Bitcoin", in Proceedings of the UbiComp Adjunct, Seattle, United States, 13-17 Sep. 2014.

[51] Zhang, Y., Wen, J., "An IoT electric business model based on the protocol of bitcoin", in Proceedings of the 18th International Conference on Intelligence in Next Generation Networks, Paris, France, 17-19 Feb. 2015.

[52] Wilkinson, S., Boshevski, T., Brandoff, J., Prestwich, J., Hall, G., Gerbes, P., Hutchins, P., Pollard, C., Buterin, V., "Storj A Peer-to-Peer Cloud Storage Network". Available online: https://storj.io/storj.pdf (Accessed on 10 April 2018)

[53] Ateniese, G., Goodrich, M. T., Lekakis, V., Papamanthou, C., Paraskevas, E., Tamassia, R., "Accountable Storage", in Proceedings of the International Conference on Applied Cryptography and Network Security, Kanazawa, Japan, 10-12 July 2017.

[54] Wilson, D., Ateniese, G., "From Pretty Good To Great: Enhancing PGP using Bitcoin and the Blockchain", in Proceedings of the International Conference on Network and System Security, New York, United States, 3-5 Nov. 2015.

[55] Gipp, B., Meuschke, N., Gernandt, A., "Decentralized Trusted Timestamping using the Crypto Currency Bitcoin", in Proceedings of the iConference, Newport Beach, United States, 24-27 Mar. 2015.

[56] Han, D., Kim, H., Jang, J. "Blockchain based smart door lock system", in Proceedings of the 2017 International Conference on Information and Communication Technology Convergence (ICTC), Jeju Island, South Korea, pp. 1165-1167, Dec. 2017.

[57] Lei, A., Cruickshank, H., Cao, Y., Asuquo, P., Ogah, C. P. A., Sun, Z. "Blockchain-Based Dynamic Key Management for Heterogeneous Intelligent Transportation Systems", in IEEE Internet of Things Journal, vol. 4, no. 6, pp. 1832-1843, Dec. 2017.

[58] Siddiqi, M., All, S. T., Sivaraman, V. , "Secure lightweight context-driven data logging for bodyworn sensing devices", in Proceedings of the 2017 5th International Symposium on Digital Forensic and Security (ISDFS), Tirgu Mures, Romania, pp. 1-6, 2017.

[59] Kshetri, N. "Can Blockchain Strengthen the Internet of Things?", in IT Professional, vol. 19, no. 4, pp. 68-72, 2017.

[60] Tanas, C., Delgado-Segura, S., Herrera-Joancomartí, J. "An Integrated Reward and Reputation Mechanism for MCS Preserving Users' Privacy", in Revised Selected Papers of the 10th International Workshop on Data Privacy Management, and Security Assurance - vol. 9481. Springer-Verlag New York, Inc.,USA, pp. 83-99, 2016.

[61] Wright, A., De Filippi, P. "Decentralized Blockchain Technology and the Rise of Lex Cryptographia", March 2015. Available online: https://ssrn.com/abstract=2580664 (Accessed on 10 April 2018)

[62] Kshetri, N. "Blockchain's roles in strengthening cybersecurity and protecting privacy", in Telecommunications Policy, vol. 41, no. 10, pp. 1027-1038, 2017.

[63] Tian, F., "An agri-food supply chain traceability system for China based on RFID & blockchain technology", in Proceedings of the 13th International Conference on Service Systems and Service Management (ICSSSM), Kunming, China, 24-26 June 2016.

[64] Huh, S., Cho, S., Kim, S., "Managing IoT devices using blockchain platform", in Proceedings of the 19th International Conference on Advanced Communication Technology (ICACT), Bongpyeong, South Korea, 19-22 Feb. 2017.

[65] Kafle, Y. R., Mahmud, K., Morsalin, S., Town, G. E., "Towards an internet of energy", in Proceedings of the IEEE International Conference on Power System Technology (POWERCON), Wollongong, Australia, 28 Sep.-1 Oct. 2016.

[66] Blanco-Novoa, O., Fernández-Caramés, T. M., Fraga-Lamas, P., Castedo, L. "An Electricity-Price Aware Open-Source Smart Socket for the Internet of Energy", in Sensors, vol. 17 (3), no. 643, pp. 1–34, March 2017.

[67] Fernández-Caramés, T.M. An Intelligent Power Outlet System for the Smart Home of the Internet of Things. in Int. J. Distrib. Sens. Netw. 2015. doi:10.1155/2015/214805.

[68] Lundqvist, T., de Blanche, A., Andersson, H. R. H. "Thing-to-thing electricity micro payments using blockchain technology", in Proceedings

of the Global Internet of Things Summit (GIoTS), Geneva, Switzerland, 6-9 June 2017.
[69] Bocek, T., Rodrigues, B. B., Strasser, T., Stiller, B. "Blockchains everywhere - a use-case of blockchains in the pharma supply-chain", in Proceedings of the IFIP/IEEE Symposium on Integrated Network and Service Management (IM), Lisbon, Portugal, 8-12 May 2017.
[70] Shae, Z., Tsai, J. J. P. "On the Design of a Blockchain Platform for Clinical Trial and Precision Medicine", in Proceedings of the IEEE 37th International Conference on Distributed Computing Systems (ICDCS), Atlanta, United States, 5-8 June 2017.
[71] Salahuddin, M. A., Al-Fuqaha, A., Guizani, M., Shuaib, K., Sallabi, F. "Softwarization of Internet of Things Infrastructure for Secure and Smart Healthcare" in Computer, vol. 50, no. 7, July 2017, pp. 74-79.
[72] Dolui, K., Datta, S. K. "Comparison of edge computing implementations: Fog computing, cloudlet and mobile edge computing", in Proceedings of the Global Internet of Things Summit (GIoTS), Geneva, Switzerland, 6-9 June 2017.
[73] Tor project official web page. Available online: https://www.torproject.org (Accessed on 10 April 2018)
[74] Park, J., Kim, K. "TM-Coin: Trustworthy management of TCB measurements in IoT", in Proceedings of the IEEE International Conference on Pervasive Computing and Communications Workshops, Kona, United States, 13-17 March 2017.
[75] ARM TrustZone official web page. Avaiable online: https://www.arm.com/products/security-on-arm/trustzone (Accessed on 10 April 2018)
[76] Biswas, K., Muthukkumarasamy, V. "Securing Smart Cities Using Blockchain Technology", in Proceedings of the IEEE 14th International Conference on Smart City, Sydney, Australia, 12-14 Dec. 2016.
[77] Sukhwani, H., Martínez, J. M., Chang, X., Trivedi, K. S., Rindos, A. "Performance Modeling of PBFT Consensus Process for Permissioned Blockchain Network (Hyperledger Fabric)", in Proceedings of the IEEE 36th Symposium on Reliable Distributed Systems (SRDS), Hong Kong, China, 26-29 Sep. 2017.
[78] Vukolić, M. "The Quest for Scalable Blockchain Fabric: Proof-of-Work vs. BFT Replication". Available online: http://www.vukolic.com/iNetSec_2015.pdf (Accessed on 10 April 2018)
[79] Bahrepour, M., Meratnia, N., Havinga, P. J. M. "Sensor fusion-based event detection in Wireless Sensor Networks", in Proceedings of the 2009 6th Annual International Mobile and Ubiquitous Systems: Networking & Services, MobiQuitous, Toronto, 2009, pp. 1-8.
[80] Anirudh, M., Thileeban, S. A., Nallathambi, D. J. "Use of honeypots for mitigating DoS attacks targeted on IoT networks", in Proceedings of the International Conference on Computer, Communication and Signal Processing (ICCCSP), Chennai, Indica, 10-11 Jan. 2017.
[81] Xu, Q., Ren, P., Song, H., Du, Q. "Security Enhancement for IoT Communications Exposed to Eavesdroppers With Uncertain Locations", in IEEE Access, vol. 4, pp. 2840-2853, June 2016.
[82] Li, X., Wang, H., Yu, Y., Qian, C. "An IoT Data Communication Framework for Authenticity and Integrity", in Proceedings of the IEEE/ACM Second International Conference on Internet-of-Things Design and Implementation (IoTDI), Pittsburgh, United States, 18-21 Apr. 2017.
[83] Yu, T., Wang, X., Shami, A. "Recursive Principal Component Analysis-Based Data Outlier Detection and Sensor Data Aggregation in IoT Systems", in IEEE Internet of Things Journal, vol. 4, no. 6, pp. 2207-2216, Dec. 2017.
[84] Raspberry Pi official web page. Available online https://www.raspberrypi.org (Accessed on 10 April 2018).
[85] Beagle boards official web page. Available online http://beagleboard.org (Accessed on 10 April 2018).
[86] K. Yeow, A. Gani, R. W. Ahmad, J. J. P. C. Rodrigues and K. Ko, "Decentralized Consensus for Edge-Centric Internet of Things: A Review, Taxonomy, and Research Issues," in IEEE Access, vol. 6, pp. 1513-1524, 2018.
[87] Liao, C.-F., Bao, S.-W., Cheng, C.-J. "On design issues and architectural styles for blockchain-driven IoT services", in Proceedings of the IEEE International Conference on Consumer Electronics - Taiwan (ICCE-TW), Taipei, Taiwan, 12-14 June 2017.
[88] IBM, "ADEPT: An IoT Practitioner Perspective", white paper, 2015.
[89] Telehash official web page. Available online: http://telehash.org (Accessed on 10 April 2018)
[90] BitTorrent official web page. Available online: http://www.bittorrent.com (Accessed on 10 April 2018)
[91] Dorri, A., Kanhere, S. S., Jurdak, R. "Blockchain in internet of things: Challenges and Solutions". Available online: https://arxiv.org/abs/1608.05187 (Accessed on 10 April 2018)
[92] Dorri, A., Kanhere, S. S., Jurdak, R. "Towards an Optimized BlockChain for IoT", in Proceedings of the IEEE/ACM Second International Conference on Internet-of-Things Design and Implementation (IoTDI), Pittsburgh, United States, 18-21 Apr. 2017.
[93] Daza, V., Di Pietro, R., Klimek, I., Signorini, M. "CONNECT: CONtextual NamE disCovery for blockchain-based services in the IoT", in Proceedings of the IEEE International Conference on Communications, Paris, France, 21-25 May 2017.
[94] Li, C., Zhang, L.-J. "A Blockchain Based New Secure Multi-Layer Network Model for Internet of Things", in Proceedings of the IEEE International Congress on Internet of Things (ICIOT), Honolulu, United States, 25-30 June 2017.
[95] Samaniego, M., Deters, R. "Blockchain as a Service for IoT", in Proceedings of the IEEE International Conference onInternet of Things (iThings) and IEEE Green Computing and Communications (GreenCom) and IEEE Cyber, Physical and Social Computing (CPSCom) and IEEE Smart Data (SmartData), Chengdu, China, 15-18 Dec. 2016.
[96] Samaniego, M., Deters, R. "Hosting Virtual IoT Resources on Edge-Hosts with Blockchain", in Proceedings of the IEEE International Conference on Computer and Information Technology (CIT), Nadi, Fiji, 8-10 Dec. 2016.
[97] Samaniego, M., Deters, R. "Internet of Smart Things - IoST: Using Blockchain and CLIPS to Make Things Autonomous", IEEE International Conference on Cognitive Computing (ICCC), Honolulu, United States, 25-30 June 2017.
[98] Stanciu, A. "Blockchain Based Distributed Control System for Edge Computing", in Proceedings of the 21st International Conference on Control Systems and Computer Science, Bucharest, Romania, 29-31 May 2017.
[99] IEC 61499 standard official web page. Available online: http://www.iec61499.de (Accessed on 10 April 2018)
[100] Docker official web page. Available online: https://www.docker.com (Accessed on 10 April 2018)
[101] Kubernetes official web page. Available online: https://kubernetes.io (Accessed on 10 April 2018)
[102] Sharma, P. K., Chen, M.-Y., Park, J.-H. "A Software Defined Fog Node based Distributed Blockchain Cloud Architecture for IoT", IEEE Access, Sep. 2017.
[103] Sharma, P. K., Singh, S., Jeong, Y.-S., Park, J.-H. "DistBlockNet: A Distributed Blockchains-Based Secure SDN Architecture for IoT Networks", IEEE Communications Magazine, vol. 55, no. 9, Sep. 2017, pp. 78-85.
[104] Li, N., Liu, D., Nepal, S. "Lightweight Mutual Authentication for IoT and Its Applications", in IEEE Transactions on Sustainable Computing, vol. 2, no. 4, pp. 359-370, Oct.-Dec. 1 2017.
[105] NIST official web page. Available online: https://www.nist.gov (Accessed on 10 April 2018)
[106] Polk, T., McKay, K., Chokhani, S. "Guidelines for the Selection and Use of Transport Layer Security (TLS) Implementations", NIST Special Publication 800-52 Revision 1, June 2005.
[107] Rivest, R. L., Shamir, A., Adleman, L. "A method for obtaining digital signatures and public-key cryptosystems", in Commun. ACM, vol. 21, no. 2, pp. 120-126, Feb. 1978.
[108] Bos, J.W., Halderman, J.A., Heninger, N., Moore, J., Naehrig, M., Wustrow E. "Elliptic Curve Cryptography in Practice". In: Christin N., Safavi-Naini R. (eds) Financial Cryptography and Data Security. FC 2014. Lecture Notes in Computer Science, vol 8437. Springer, Berlin, Heidelberg.
[109] Kleinjung, T., Aoki, K., Franke, J., Lenstra, A. K., Thomé, E., Bos, J. W., Gaudry, P., Kruppa, A., Montgomery, P. L., Osvik, D. A., te Riele, H. J. J., Timofeev, A., Zimmermann, P. "Factorization of a 768-Bit RSA Modulus", in Proceedings of the 30th annual conference on Advances in cryptology, Santa Barbara, United States, 15-19 Aug. 2010.
[110] Pellegrini, A., Bertacco, V., Austin, T. "Fault-based attack of RSA authentication", in Proceedings of the Design, Automation & Test in Europe Conference & Exhibition, Dresden, Germany, 8-12 Mar. 2010.
[111] Levi, A., Savas, E. "Performance evaluation of public-key cryptosystem operations in WTLS protocol", in Proceedings of the Eighth IEEE Symposium on Computers and Communications, Kemer-Antalya, Turkey, 30 June-3 July 2003.
[112] Habib, M., Mehmood, T., Ullah, F., Ibrahim, M. "Performance of WiMAX Security Algorithm (The Comparative Study of RSA Encryp-

tion Algorithm with ECC Encryption Algorithm)", in Proceedings of the 2009 International Conference on Computer Technology and Development, Kota Kinabalu, Malaysia, 13-15 Nov. 2009.
[113] Gura, N., Patel, A., Wander, A., Eberle, H., Shantz, S. C. "Comparing Elliptic Curve Cryptography and RSA on 8-bit CPUs", in Proceedings of the International Workshop on Cryptographic Hardware and Embedded Systems, Cambridge, United States, 11-13 Aug. 2004.
[114] Savari, M., Montazerolzohour, M., Thiam, Y. E. "Comparison of ECC and RSA algorithm in multipurpose smart card application", in Proceedings of the International Conference on Cyber Security, Cyber Warfare and Digital Forensic, Kuala Lumpur, Malaysia, 26-28 June 2012.
[115] Bafandehkar, M., Yasin, S. M., Mahmod, R., Hanapi, Z. M. "Comparison of ECC and RSA Algorithm in Resource Constrained Devices", in Proceedings of the International Conference on IT Convergence and Security, Macau, China, 16-18 Dec. 2013.
[116] Wander, A. S., Gura, N., Eberle, H., Gupta, V., Shantz, S. C. "Energy analysis of public-key cryptography for wireless sensor networks", in Proceedings of the Third IEEE International Conference on Pervasive Computing and Communications, Kauai Island, United States, 8-12 Mar. 2005.
[117] Noroozi, E., Kadivar, J., Shafiee, S. H. "Energy analysis for wireless sensor networks", in Proceedings of the 2nd International Conference on Mechanical and Electronics Engineering, Kyoto, Japan, 1-3 Aug. 2010.
[118] de Oliveira, P. R., Feltrim, V. D., Fondazzi Martimiano, L. A., Marcal Zanoni, G. B. "Energy Consumption Analysis of the Cryptographic Key Generation Process of RSA and ECC Algorithms in Embedded Systems", IEEE Latin America Transactions, vol. 6, no. 6, pp. 1141-1148, Sep. 2014.
[119] Goyal, T. K., Sahula, V. "Lightweight security algorithm for low power IoT devices", in Proceedings of the 2016 International Conference on Advances in Computing, Communications and Informatics, Jaipur, India, 21-24 Sep. 2016.
[120] Koblitz, N., Menezes, A. "A Riddle Wrapped in an Enigma", in IEEE Security & Privacy, vol 14, no. 6, pp. 34-42, Dec. 2016.
[121] Rogaway, P., Shrimpton, T. "Cryptographic hash-function basics: Definitions, implications, and separations for preimage resistance, second-preimage resistance, and collision resistance", in: Bimal Roy, Willi Meier (Eds.), Proceedings of the 11th Fast Software Encryption, FSE'04, in Lecture Notes in Computer Science, vol. 3017, Springer Verlag, pp. 371–388, 2004.
[122] Ometov, A., Masek, P. Malina, L., Florea, R., Hosek, J., Andreev, S., Hajny, J., Niutanen, J., Koucheryavy, Y. "Feasibility characterization of cryptographic primitives for constrained (wearable) IoT devices", in Proceedings of the IEEE International Conference on Pervasive Computing and Communication Workshops, Sydney, Australia, 14-18 Mar. 2016,
[123] Feldhofer, M., Rechberger, C. "A Case Against Currently Used Hash Functions in RFID Protocols", in Proceedings of On the Move to Meaningful Internet Systems Workshops, Montpellier, France, 29 Oct.-3 Nov. 2006.
[124] Degnan, B., Durgin, G., Maeda, S. "On the Simon Cipher 4-block key schedule as a hash", in Proceedings of IEEE International Conference on RFID, Phoenix, United States, 9-11 May 2017.
[125] Haber, S., Stornetta, W. S. "How to time-stamp a digital document", in Journal of Cryptology, vol. 3, no. 2, pp. 99-111, Jan. 1991.
[126] Douceur, J. R. "The Sybil Attack", in Proceedings of the 1st International Workshop on Peer-to-Peer Systems (IPTPS), Cambridge, United States, 7-8 March 2002.
[127] Sato, M., Matsuo, S. "Long-Term Public Blockchain: Resilience against Compromise of Underlying Cryptography", in Proceedings of the IEEE European Symposium on Security and Privacy Workshops, Vancouver, Canada, 31 July-3 Ago. 2017.
[128] Takura, A., Ono, S., Naito, S. "A secure and trusted time stamping authority", in Proceedings of the Internet Workshop, Osaka, Japan, 18-20 Feb. 1999.
[129] Peercoin official web page. Available online: https://peercoin.net (Accessed on 10 April 2018)
[130] DPOS description on Bitshares. Available on: http://docs.bitshares.org/bitshares/dpos.html (Accessed on 10 April 2018)
[131] Larimer, D. "Transactions as proof-of-stake". Available online: https://bravenewcoin.com/assets/Uploads/TransactionsAsProofOfStake10.pdf (Accessed on 10 April 2018)
[132] Bentov, I., Lee, C., Mizrahi, A., Rosenfeld, M. "Proof of activity: Extending bitcoin's proof of work via proof of stake", in Proceedings of the 9th Workshop on the Economics of Networks, Systems and Computation, Austin, United States, 16 June 2014.
[133] Ren, L. "Proof of Stake Velocity: Building the Social Currency of the Digital Age". Available online: https://www.reddcoin.com/papers/PoSV.pdf (Accessed on 10 April 2018).
[134] Reddcoin official web page. Available online: www.reddcoin.com (Accessed on 10 April 2018).
[135] Castro, M., Liskov, B. "Practical Byzantine fault tolerance", in Proceedings of the Third Symposium on Operating Systems Design and Implementation, New Orleans, United States, Feb. 1999.
[136] Schwartz, D., Youngs, N., Britto, A. "The ripple protocol consensus algorithm", white paper, Ripple Labs, 2014.
[137] Mazieres, D. "The stellar consensus protocol: A federated model for internet-level consensus". Available online: https://www.stellar.org/papers/stellar-consensus-protocol.pdf (Accessed on 10 April 2018).
[138] Copeland, C., Zhong, H. "Tangaroa: A Byzantine fault tolerant raft", Available online: http://www.scs.stanford.edu/14au-cs244b/labs/projects/copeland_zhong.pdf (Accessed on 10 April 2018)
[139] Ongaro, D., Ousterhout, J. "In Search of an Understandable Consensus Algorithm", in Proceedings of USENIX Annual Technical Conference, Philadelphia, United States, 19–20 June, 2014.
[140] Cachin, C., Schubert, S., Vukolić, M. "Non-determinism in Byzantine fault-tolerant replication", in Proceedings of the International Conference on Principles of Distributed Systems (OPODIS 2016), Madrid, Spain, 13-16 Dec. 2016.
[141] Kwon, J. "Tendermint: Consensus without mining (v0.6)". Available online: https://tendermint.com/static/docs/tendermint.pdf (Accessed on 10 April 2018)
[142] Eyal, I., Gencer, A. E., Sirer, E. G., Van Renesse, R. "Bitcoin-NG: A Scalable Blockchain Protocol", in Proceedings of the 13th USENIX Symposium on Networked Systems Design and Implementation, Santa Clara, United States, 16-18 Mar. 2016.
[143] Borge, M., Kokoris-Kogias, E., Jovanovic, P., Gasser, L., Gailly, N., Ford, B. "Proof-of-Personhood: Redemocratizing Permissionless Cryptocurrencies", in Proceedings of the IEEE European Symposium on Security and Privacy Workshops, Paris, France, 26-28 Apr. 2017.
[144] Rivest, R. L., Shamir, A., Tauman, Y. "How to Leak a Secret", in Proceedings of the 7th International Conference on the Theory and Application of Cryptology and Information Security, Gold Coast, Australia, 9-13 Dec. 2001, pp. 552-565.
[145] Syta, E., Tamas, I., Visher, D., Wolinsky, D., Gasser, L., Gailly, N., Ford, B. "Keeping Authorities "Honest or Bust" with Decentralized Witness Cosigning", in Proceedings of the 37th IEEE Symposium on Security and Privacy, San José, United States, 23-25 May 2016.
[146] PoI project official web page. Available online: http://proofofindividuality.online (Accessed on 10 April 2018)
[147] Kambourakis, G., Kolias, C., Stavrou, A. "The Mirai botnet and the IoT Zombie Armies", in Proceedings of the IEEE Military Communications Conference (MILCOM), Baltimore, United States, 23-25 Oct. 2017.
[148] Chandra, H., Anggadjaja, E., Wijaya, P. S., Gunawan, E. "Internet of Things: Over-the-Air (OTA) firmware update in Lightweight mesh network protocol for smart urban development", in Proceedings of the 22nd Asia-Pacific Conference on Communications (APCC), Yogyakarta, Indonesia, 25-27 Aug. 2016.
[149] Boudguiga, A., Bouzerna, N., Granboulan, L., Olivereau, A., Quesnel, F., Roger, A., Sirdey, R. "Towards Better Availability and Accountability for IoT Updates by Means of a Blockchain", in Proceedings of the IEEE European Symposium on Security and Privacy Workshops, Paris, France, 26-28 Apr. 2017.
[150] Meher, R. "The Internet of Money". Available online: https://docs.google.com/document/d/1Bc-kZXROTeMzG6AvH7rrTrUy24UwHoEcgiL7ALHMO0A/pub (Accessed on 10 April 2018)
[151] Fraga-Lamas, P., Noceda-Davila, D., Fernández-Caramés, T.M., Díaz-Bouza, M., Vilar-Montesinos, M. "Smart Pipe System for a Shipyard 4.0", in Sensors, vol. 16 (12), no. 2186, pp. 1–43, December 2016.
[152] Fraga-Lamas, P., Fernández-Caramés, T.M., Noceda-Davila, D., Vilar-Montesinos, M. "RSS Stabilization Techniques for a Real-Time Passive UHF RFID Pipe Monitoring System for Smart Shipyards", in Proceedings of the 2017 IEEE International Conference on RFID (IEEE RFID 2017), Phoenix, AZ, USA, 9–11 May 2017.
[153] Fraga-Lamas, P., Fernández-Caramés, T.M., Noceda-Davila, D., Díaz-Bouza, M., Vilar-Montesinos, M., Pena-Agras J. D., Castedo, L. "Enabling Automatic Event Detection for the Pipe Workshop of the Shipyard

4.0", in Proceedings of the 2017 56th FITCE Congress, Madrid, Spain, 14-16 September 2017, pp. 20-27.
[154] Barro-Torres, S.J., Fernández-Caramés, T.M., González-López, M., Escudero-Cascón, C.J. "Maritime Freight Container Management System Using RFID", in Proceedings of the Third International EURASIP Workshop on RFID Technology, La Manga del Mar Menor, Spain, 6–7 September 2010.
[155] Hernández-Rojas, D.L., Fernández-Caramés, T.M., Fraga-Lamas, P., Escudero, C.J. "Design and Practical Evaluation of a Family of Lightweight Protocols for Heterogeneous Sensing through BLE Beacons in IoT Telemetry Applications", in Sensors, vol. 18 (1), no. 57, pp. 1–33, December 2017.
[156] Fraga-Lamas, P., Castedo-Ribas, L., Morales-Méndez, A., Camas-Albar, J.M. "Evolving military broadband wireless communication systems: WiMAX, LTE and WLAN", in Proceedings of the International Conference on Military Communications and Information Systems (ICMCIS), Brussels, Belgium, 23–24 May 2016; pp. 1–8.
[157] Fraga-Lamas, P., Rodríguez-Piñeiro, J., García-Naya, J.A., Castedo, L. "Unleashing the potential of LTE for next generation railway communications", in Proceedings of the 8th International Workshop on Communication Technologies for Vehicles (Nets4Cars/Nets4Trains/Nets4Aircraft 2015), Sousse, Tunisia, 6–8 May 2015; Lecture Notes in Computer Science; Springer: Berlin/Heidelberg, Germany, 2015; Volume 9066, pp. 153–164.
[158] Fraga-Lamas, P. "Enabling Technologies and Cyber-Physical Systems for Mission-Critical Scenarios", PhD dissertation. University of A Coruña, 2017.
[159] Meiklejohn, S., Pomarole, M., Jordan, G., Levchenko, K., McCoy, D., Voelker, G. M., Savage, S. "A fistful of bitcoins: Characterizing payments among men with no names", Communications of the ACM, vol. 59, no. 4, pp. 86-93, April 2016.
[160] Möser, M., Böhme, R., Breuker, D. "An inquiry into money laundering tools in the Bitcoin ecosystem", in Proceedings of the APWG eCrime Researchers Summit, San Francisco, United States, 17-18 Sep. 2013.
[161] Fabiano, N. "The Internet of Things ecosystem: The blockchain and privacy issues. The challenge for a global privacy standard", in Proceedings of the International Conference on Internet of Things for the Global Community (IoTGC), Funchal, Portugal, 10-13 July 2017, pp. 1-7.
[162] Kravitz, D. W., Cooper, J. "Securing user identity and transactions symbiotically: IoT meets blockchain", in Proceedings of the Global Internet of Things Summit (GIoTS), Geneva, Switzerland, 6-9 June 2017.
[163] Hashemi, S. H., Faghri, F., Rausch, P., Campbell, R. H. "World of Empowered IoT Users", in Proceedings of the IEEE First International Conference on Internet-of-Things Design and Implementation (IoTDI), Berlin, Germany, 4-8 Apr. 2016, pp. 13-24.
[164] Fraga-Lamas, P., Fernández-Caramés, T. M. "Reverse Engineering the Communications Protocol of an RFID Public Transportation Card", in Proceedings of the 2017 IEEE International Conference on RFID (IEEE RFID 2017), Phoenix, AZ, USA, 9–11 May 2017; pp. 30–35.
[165] Fernández-Caramés, T. M., Fraga-Lamas, P., Suárez-Albela, M., Castedo, L. "Reverse Engineering and Security Evaluation of Commercial Tags for RFID-Based IoT Applications", in Sensors, vol. 17 (1), no. 28, pp. 1–31, December 2016.
[166] Fernández-Caramés, T. M., Fraga-Lamas, P., Suárez-Albela, M., Castedo, L. "A Methodology for Evaluating Security in Commercial RFID Systems, Radio Frequency Identification", in Radio Frequency Identification, 1st ed.; Crepaldi, P. C., Pimenta, T. C., Eds.; INTECH: Rijeka, Croatia, 2017.
[167] Li, Z., Braun, T. "Passively Track WiFi Users With an Enhanced Particle Filter Using Power-Based Ranging", in IEEE Transactions on Wireless Communications, vol. 16, no. 11, pp. 7305-7318, Nov. 2017.
[168] Luo, C., Cheng, L., Chan, M. C., Gu, Y., Li, J., Ming, Z. "Pallas: Self-Bootstrapping Fine-Grained Passive Indoor Localization Using WiFi Monitors", in IEEE Transactions on Mobile Computing, vol. 16, no. 2, pp. 466-481, Feb. 2017.
[169] Multichain white paper. Available online: https://www.multichain.com/download/MultiChain-White-Paper.pdf (Accessed on 10 April 2018)
[170] Danezis, G., Serjantov, A. "Statistical Disclosure or Intersection Attacks on Anonymity Systems", in Proceedings of the 6th International Workshop on Information Hiding, Toronto, Canada, 23-25 May, 2004, pp. 293-308.
[171] Bonneau, J., Narayanan, A., Miller, A., Clark, J., Kroll, J. A., Felten, E. W., Christin, N., Safavi-Naini, R. "Mixcoin: Anonymity for Bitcoin with Accountable Mixes", in Proceedings of the 18th International Conference on Financial Cryptography and Data Security, Christ Church, Barbados, 3-7 March, 2014, pp. 486-504.
[172] Valenta, L., Rowan, B. "Blindcoin: Blinded, Accountable Mixes for Bitcoin", International Workshops on BITCOIN, WAHC, and Wearable, San Juan, Puerto Rico, 30 Jan., 2015, pp. 112-126.
[173] Zerocoin official web page. Available online: http://zerocoin.org (Accessed on 10 April 2018).
[174] Zerocash official web page. Available online: http://zerocash-project.org (Accessed on 10 April 2018).
[175] Zcash official web page. Available online: https://z.cash (Accessed on 10 April 2018).
[176] Schukat, M., Flood, P. "Zero-knowledge proofs in M2M communication", in Proceedings of the 25th IET Irish Signals & Systems Conference and China-Ireland International Conference on Information and Communications Technologies, Limerick, Ireland, 26-27 June 2014.
[177] Peng, K. "Attack against a batch zero-knowledge proof system", in IET Information Security, vol. 6, no. 1, pp. 1-5, March 2012.
[178] Bytecoin's official web page. Available online: https://bytecoin.org (Accessed on 10 April 2018).
[179] Monero's official web page. Available online: https://getmonero.org (Accessed on 10 April 2018).
[180] CryptoNote's official web page. Available online: https://cryptonote.org (Accessed on 10 April 2018).
[181] Moore, C., O'Neill, M., O'Sullivan, E., Doröz, Y., Sunar, B. "Practical homomorphic encryption: A survey", in Proceedings of the IEEE International Symposium on Circuits and Systems (ISCAS), Melbourne, Australia, 1-5 June 2014.
[182] Hayouni, H., Hamdi, M. "Secure data aggregation with homomorphic primitives in wireless sensor networks: A critical survey and open research issues", in Proceedings of the IEEE 13th International Conference on Networking, Sensing, and Control (ICNSC), Mexico City, Mexico, 28-30 Apr. 2016.
[183] França, B. F. "Homomorphic Mini-blockchain Scheme", April 2015. Available online: http://cryptonite.info/files/HMBC.pdf (Accessed on 10 April 2018)
[184] Lukianov, D. "Compact Confidential Transactions for Bitcoin", December 2015. Available online: http://voxelsoft.com/dev/cct.pdf (Accessed on 10 April 2018).
[185] Jabir, R. M., Khanji, S. I. R., Ahmad, L. A., Alfandi, O., Said, H. "Analysis of cloud computing attacks and countermeasures", in Proceedings of the 18th International Conference on Advanced Communication Technology (ICACT), Pyeongchang, South Korea, 31 Jan.-3 Feb. 2016.
[186] Atya, A. O. F., Qian, Z., Krishnamurthy, S. V., Porta, T. L., McDaniel, P., Marvel, L. "Malicious co-residency on the cloud: Attacks and defense", in Proceedings of the IEEE Conference on Computer Communications, Atlanta, United States, 1-4 May 2017.
[187] CONIKS official web page. Available online: https://coniks.cs.princeton.edu (Accessed on 10 April 2018).
[188] Chen, T. M., Abu-Nimeh, S. "Lessons from Stuxnet" in Computer, vol. 44, no. 4, pp. 91-93, Apr. 2011.
[189] Google's Certificate Transparency official web page. Available online: https://www.certificate-transparency.org (Accessed on 10 April 2018).
[190] Liu, B., Yu, X. L., Chen, S., Xu, X., Zhu, L. "Blockchain Based Data Integrity Service Framework for IoT Data", in Proceedings of the IEEE International Conference on Web Services, Honolulu, United States, 25-30 June 2017.
[191] Zhou, Z., Xie, M., Zhu, T., Xu, W., Yi, P., Huang, Z., Zhang, Q., Xiao, S. "EEP2P: An energy-efficient and economy-efficient P2P network protocol", in Proceedings of the International Green Computing Conference, Dallas, United States, 3-5 Nov. 2014.
[192] Sharifi, L., Rameshan, N., Freitag, F., Veiga, L. "Energy Efficiency Dilemma: P2P-cloud vs. Datacenter", in Proceedings of the IEEE 6th International Conference on Cloud Computing Technology and Science, Singapore, Singapore, 15-18 Dec. 2014.
[193] Zhang, P., Helvik, B. E. "Towards green P2P: Analysis of energy consumption in P2P and approaches to control", in Proceedings of the International Conference on High Performance Computing & Simulation (HPCS), Madrid, Spain, 2-6 July 2012.
[194] Miyake, S., Bandai, M. "Energy-Efficient Mobile P2P Communications Based on Context Awareness", in Proceedings of the IEEE 27th International Conference on Advanced Information Networking and Applications (AINA), Barcelona, Spain, 25-28 Mar. 2013.
[195] Liao, C. C., Cheng, S. M., Domb, M. "On Designing Energy Efficient Wi-Fi P2P Connections for Internet of Things", in Proceedings of the IEEE

85th Vehicular Technology Conference (VTC Spring), Sydney, Australia, 4-7 June 2017.
[196] Ball, M., Rosen, A., Sabin, M., Vasudevan, P. N. "Proofs of Useful Work". Available online: https://eprint.iacr.org/2017/203.pdf (Accessed on 10 April 2018)
[197] Gridcoin's official web page. Available online: http://gridcoin.us (Accessed on 10 April 2018)
[198] Primecoin's official web page. Available online: http://www.primecoin.org (Accessed on 10 April 2018)
[199] Dziembowski, S., Faust, S., Kolmogorov, V., Pietrzak, K. "Proofs of Space", in Proceedings of the 35th Annual Cryptology Conference on Advances in Cryptology, Santa Barbara, United States, 16-20 Aug. 2015, pp. 585-605.
[200] Burst-coin official web page. Available online: https://www.burst-coin.org (Accessed on 10 April 2018)
[201] Bruce, J. D. "The Mini-Blockchain Scheme". Available online: https://www.weusecoins.com/assets/pdf/library/The\%20Mini-Blockchain\%20Scheme.pdf (Accessed on 10 April 2018)
[202] Original Scrypt function for Tarsnap. Available online: http://www.tarsnap.com/scrypt.html (Accessed on 10 April 2018)
[203] X11 official documentation for Dash. Available online: https://dashpay.atlassian.net/wiki/spaces/DOC/pages/1146918/X11 (Accessed on 10 April 2018)
[204] Aumasson, J.-P., Henzen, L., Meier, W., Phan, R. C.-W. "SHA-3 Proposal BLAKE, submission to NIST". Available online: http://131002.net/blake/ (accessed on 10 April 2018)
[205] Myriad. Available online: http://myriadcoin.org (Accessed on 10 April 2018)
[206] Courtois, N. T., Emirdag, P., Nagy, D. A. "Could Bitcoin transactions be 100x faster?", in Proceedings of the 11th International Conference on Security and Cryptography (SECRYPT), Vienna, Austria, 28-30 Aug. 2014.
[207] VISA claims about the number of transactions handled by VisaNet. Available online: https://usa.visa.com/run-your-business/small-business-tools/retail.html (Accessed on 10 April 2018)
[208] Bedford Taylor, M. "The Evolution of Bitcoin Hardware" in Computer, vol. 50, no. 9, pp. 58-66, 22 Sep. 2017.
[209] Birrell, E., Schneider, F. B. "Federated Identity Management Systems: A Privacy-Based Characterization", in IEEE Security & Privacy, vol. 11, no. 5, pp. 36-48, Sept.-Oct. 2013.

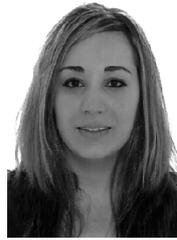
PAULA FRAGA-LAMAS (M'17) received the M.Sc. degree in Computer Science in 2008 from University of A Coruña (UDC) and the M.Sc. and Ph.D. degrees in the joint program Mobile Network Information and Communication Technologies from five Spanish universities: University of the Basque Country, University of Cantabria, University of Zaragoza, University of Oviedo and University of A Coruña, in 2011 and 2017, respectively. Since 2009, she has been working with the Group of Electronic Technology and Communications (GTEC) in the Department of Computer Engineering (UDC). She is co-author of more than thirty peer-reviewed indexed journals, international conferences and book chapters. Her current research interests include wireless communications in mission-critical scenarios, Industry 4.0, Internet of Things (IoT), Augmented Reality (AR), blockchain, RFID and Cyber-Physical systems (CPS). She has also been participating in more than twenty research projects funded by the regional and national government as well as R&D contracts with private companies.

...

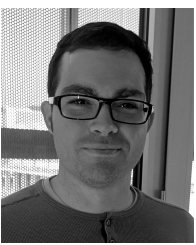
TIAGO M. FERNÁNDEZ-CARAMÉS (S'08-M'12-SM'15) received his MSc degree and PhD degrees in Computer Science in 2005 and 2011 from University of A Coruña, Spain. Since 2005 he has worked as a researcher and professor for the Department of Computer Engineering of the University of A Coruña inside the Group of Electronic Technology and Communications (GTEC). His current research interests include IoT systems, RFID, wireless sensor networks, Industry 4.0, blockchain and augmented reality.